\documentclass[preprint,3p,sort&compress]{elsarticle}

\usepackage[usenames,dvipsnames,svgnames,table]{xcolor}
\usepackage{graphicx}
\usepackage{dcolumn}
\usepackage{bm}
\usepackage[colorlinks]{hyperref}
\makeatletter

\def\ps@pprintTitle{%
 \let\@oddhead\@empty
 \let\@evenhead\@empty
 \def\@oddfoot{\centerline{\thepage}}%
 \let\@evenfoot\@oddfoot}
\makeatother

\usepackage{amsmath}	
\usepackage{amssymb}
\usepackage{xcolor}
\usepackage{graphicx}
\usepackage{caption}
\usepackage{subcaption}
\usepackage[T1]{fontenc} 
\usepackage[utf8x]{inputenc} 
\def\mbY{\ensuremath{m_{b,Y}}}

\def\mb{\ensuremath{m_{b}}}
\def\as{\ensuremath{\alpha_s}}
\def\al{\ensuremath{\alpha}}
\def\alF{\ensuremath{\alpha_F}}
\def\vF{\ensuremath{v_F}}
\def\deltaF{\ensuremath{\tilde\delta}}
\def\Veff{\ensuremath{V_{\mathrm{eff}}}}
\def\Vtree{\ensuremath{V_{\mathrm{tree}}}}
\def\veff{\ensuremath{v_{\mathrm{eff}}}}
\def\vtree{\ensuremath{v_{\mathrm{tree}}}}

\def\GeV{\ensuremath{\mathrm{GeV}}}

\def\asopi{\ensuremath{a_s}}
\def\alopi{\ensuremath{a}}

\def\effas{\ensuremath{{\alpha^\prime_s}}}
\def\effal{\ensuremath{{\alpha'}}}

\def\effasopi{\ensuremath{{a^\prime_s}}}
\def\effalopi{\ensuremath{{a^\prime}}}

\def\alFopi{\ensuremath{a_F}}

\def\effmb{\ensuremath{\underline m_b}}
\def\LL{\ensuremath{\ln\frac{\mu^2}{\mu_0^2}}}
\def\LLsq{\ensuremath{\ln^2\frac{\mu^2}{\mu_0^2}}}
\def\LMb{\ensuremath{L_b}}
\def\LMW{\ensuremath{L_W}}
\def\LMZ{\ensuremath{L_Z}}
\def\LMT{\ensuremath{L_t}}
\def\LMWH{\ensuremath{L_{WH}}}
\def\LMWZ{\ensuremath{L_{WZ}}}
\def\LMWT{\ensuremath{L_{Wt}}}

\def\DeltaMT{\ensuremath{\Delta M_t}}
\def\DeltaMW{\ensuremath{\Delta M_W}}
\def\DeltaMZ{\ensuremath{\Delta M_Z}}
\def\DeltaMH{\ensuremath{\Delta M_H}}

\def\LLMb{\ensuremath{L_b^2}}
\newcommand{\XX}[2]{ \ensuremath{X^{(#1)}_{#2}}} 
\newcommand{\uXX}[2]{ \ensuremath{\tilde{X}^{(#1)}_{#2}}} 
\def\MSbar{\ensuremath{\overline{\mathrm{MS}}}}

\journal{Physics Letters B}

\begin{document}

\begin{frontmatter}

\title{
\vskip-1.5cm{\baselineskip14pt\rm
	\centerline{\normalsize DESY 16-234\hfill ISSN 0418-9833}
	\centerline{\normalsize December 2016\hfill}}
	\vskip1.5cm
On the $b$-quark running mass in QCD and the SM}


\author[a,b]{A.V.~Bednyakov}
\ead{alexander.bednyakov@jinr.ru}
\author[c]{B.A.~Kniehl}
\ead{kniehl@desy.de}
\author[c]{A.F.~Pikelner\footnote{On leave of absence from Joint Institute for
Nuclear Research, 141980 Dubna, Russia.}}
\ead{andrey.pikelner@desy.de}
\author[c]{O.L.~Veretin}
\ead{oleg.veretin@desy.de}

\address[a]{Joint Institute for Nuclear Research, 141980 Dubna, Russia}
\address[b]{Dubna State University, 141982, Dubna, Russia}
\address[c]{II Institut f\"ur Theoretische Physik,  Universit\"at Hamburg, \\ Luruper Chaussee 149, 22761 Hamburg, Germany}

\begin{abstract}
	We consider electroweak corrections to the relation between the running $\MSbar$ mass $m_b$ of the $b$ quark 
	in the five-flavor QCD$\times$QED effective theory and its counterpart in the Standard Model (SM). 
	As a bridge between the two parameters, we use the pole mass $M_b$ of the $b$ quark, which can be calculated in both models. 	
	The running mass 
	is not a fundamental parameter of the SM Lagrangian, but the product of the running Yukawa coupling $y_b$ and the
	Higgs vacuum expectation value. Since there exist different prescriptions to define the latter, the relations 
	considered in the paper involve a certain amount of freedom.  
	All the definitions can be related to each other in perturbation theory.
	Nevertheless, we argue in favor of a certain gauge-independent prescription and provide a relation which can 
	be directly used to deduce the value of the Yukawa coupling of the $b$ quark at the electroweak scale from 
	its effective QCD running mass. This approach allows one to resum 
	large logarithms $\ln(m_b/M_t)$ 
	systematically. Numerical analysis shows that, indeed, the corrections to the proposed relation
	are much smaller than those between $y_b$ and $M_b$. 
\end{abstract}

\end{frontmatter}

\section{Introduction}
\setcounter{footnote}{1}
Since its theoretical prediction in 1973 \cite{Kobayashi:1973fv} 
and experimental discovery in 1977 \cite{Herb:1977ek}, the bottom quark has been serving as a unique probe in studying 
	various aspects of modern particle physics.
	Special B factories with appropriate detectors, e.g., BaBar and Belle, were built and established CP violation in mesons involving $b$ quarks (see Ref.~\cite{Bevan:2014iga}
	for a comprehensive review). 
	Moreover, a dedicated LHC experiment, LHCb, aimed to further improve our knowledge of the origin of CP violation, was set up at CERN.

	Another important point is New Physics (NP), which can manifest itself in rare (flavor-violating) decays of B mesons.
	The latter, being suppressed in the Standard Model (SM), are very sensitive to contributions from beyond-the-SM (BSM) 
	theories and, at the moment, play a very important role in constraining parameter spaces of NP scenarios. 
	Finally, both the SM Higgs boson and the top quark prefer to decay into $b$ quarks.
	Due to this, the properties of the bottom quark deserve to be carefully investigated and analyzed. 

	Theoretical descriptions of the above-mentioned processes usually involve several different scales, ranging from the long-distance 
	(QCD) scale $\Lambda_\mathrm{QCD}\sim 10^{-1}$~GeV over the electroweak (EW) scale $M_Z\sim 10^2$~GeV up to NP scales $\Lambda_\mathrm{NP}\gtrsim 10^3$~GeV (see, e.g., Ref.~\cite{Fleischer:2015mla}).
	Since the scales are well separated, one can make use of effective field theories (EFTs) (see, e.g, Ref.~\cite{Pich:1998xt,Grozin:2004yc})
	and ``factorize'' physics relevant to strong (long-distance) QCD dynamics, which are difficult to calculate, yet not very interesting 
	from the fundamental point of view, from short-distance effects due to EW or NP interactions. 

	In the SM, the $b$-quark mass is generated via the Higgs mechanism due to the Yukawa interaction of the $b$ quark with the Higgs field 
	condensate $|\langle \Phi\rangle|^2 \equiv v^2/2$. 
	However, a well-known fact is that strong interactions prevent quarks from being observed as ``free'' particles, so that the notion of
	physical mass is ill-defined in this case (see, e.g., Ref.~\cite{Bigi:1994em}). 
	In such a situation, one can choose a convenient mass definition depending on the problem considered. 
	Among these definitions are the pole mass $M_b$ \cite{Tarrach:1980up}, the running mass $m_b$ and various ``threshold'' masses, such as the 1S mass $m_b^{\mathrm{1S}}$ \cite{Hoang:1998ng}, the 
	potential-subtracted (PS) mass $m_b^{\mathrm{PS}}$ \cite{Beneke:1998rk},
        the renormalon-subtracted (RS) mass $m_b^{\mathrm{RS}}$ \cite{Pineda:2001zq}, etc. (see Ref.~\cite{ElKhadra:2002wp} for a review).
	In principle, all these mass parameters can be related to each other and to the Yukawa coupling $y_b$. 
	The latter enters the fundamental SM Lagrangian and is definitely very important for precise theoretical predictions both of $B$-hadron 
	decay properties and the SM Higgs decay width.

	It is worth mentioning that the direct measurement of the $b$-quark Yukawa coupling 
	is very challenging \cite{Peskin:2012we}. 
	The estimated uncertainty for the LHC is about 20\%, and a linear collider is needed to
	reduce it by an order of magnitude \cite{Peskin:2012we,Klute:2013cx}. 

	In this paper, we address the problem of the extraction of the running\footnote{In what follows, we employ modified minimal subtraction ($\MSbar$) scheme.}
		Yukawa coupling $y_b(\mu)$ in the SM from a $b$-quark mass parameter.
		Our goal is to improve a well-established relation between $y_b(\mu)$ and $M_b$ \cite{Hempfling:1994ar,Kniehl:2004hfa,Kniehl:2014yia} at the two-loop order\footnote{In fact, QCD corrections for the case of a single heavy quark 
		are known through the four-loop level \cite{Marquard:2015qpa}. However, only three-loop terms \cite{Bekavac:2007tk,Bekavac:2009gz} are known for the case when one additional massive quark is in the spectrum (see discussion below).} of perturbation theory (PT)
	by trading the ill-defined pole mass for a short-distance mass parameter $\effmb(\mu)\equiv m_b^{(5)}(\mu)$ defined in the $\MSbar$ renormalization scheme.
	The latter plays the role of an independent Lagrangian parameter of the QCD EFT with five active quark flavors, which is valid significantly below the EW
	scale. This mass parameter is not sensitive to long-distance physics and can be extracted from experiment with much higher precision. 
	The  value of $\mu_b \equiv \effmb(\effmb)=4.18\pm0.03$~GeV quoted by the Particle Data Group (PDG) \cite{Agashe:2014kda} will be used here as an input for the determination of $y_b(\mu)$.

	This paper is organized as follows. In section~\ref{sec:running_mass_in_SM}, we consider various definitions of running quark mass,
	all of which are proportional to the running Yukawa couping $y_b(\mu)$, but differ from each other by the treatment of the vacuum expectation value (vev) $v$.
	In section~\ref{sec:matching}, we describe our procedure, which allows us to obtain the relation between $y_b(\mu)$ and $\effmb(\mu)$ 
	at a certain (matching) scale $\mu$. Section~\ref{sec:numerics} is devoted to our numerical analysis of different corrections to this relation
	and the comparison of the latter with the corresponding contributions to the $y_b$-$M_b$ relation.
	Finally, in section~\ref{sec:conclusions}, some discussions and conclusions can be found. In 
	\ref{sec:eft_RGE}, three-loop renormalization group equations (RGEs)
	for the QCD$\times$QED effective theory, utilized to relate the values of $\mu_b$ and $\effmb(\mu)$, are presented.
	In 
	\ref{sec:zm_nm}, we also include the three-loop relation between the running mass $\mb^{(6)}(\mu)$ and $M_b$ in six-flavor 
	QCD, with the account of the heavy top quark. The latter can be used to extract the running mass of the $b$ quark in the SM from its given pole mass. 

\section{\label{sec:running_mass_in_SM}Running quark masses in the SM}
	
	Fermion masses are not fundamental parameters of the SM Lagrangian, but are induced in the spontaneously broken phase and are
	proportional to the corresponding Yukawa couplings: 
\begin{align}
	m_f & =  \frac{y_f v}{\sqrt 2}, 
	\label{eq:mf_yf}
\end{align}
with $f=l,q$ denoting leptons and quarks, respectively. The Higgs field expectation value $v$ corresponds to a minimum 
of the full effective potential $\Veff(\phi)$ \cite{Coleman:1973jx,Jackiw:1974cv} for the neutral component of the Higgs doublet $\phi$, 
\begin{align}
	\Veff(\phi) = \Vtree(\phi) +  \Delta V(\phi), \qquad \Vtree(\phi) = -m_\Phi^2 \frac{\phi^2}{2} + \lambda \frac{\phi^4}{4}. 
	\label{eq:eff_pot}
\end{align}
The vev $v$ is a nonperturbative quantity that should be a 
function of fundamental SM Lagrangian parameters, i.e.,
dimensionless gauge and Yukawa couplings together with the Higgs self-coupling $\lambda$ and mass $m_\Phi$ from the 
tree-level Higgs potential $\Vtree$. 
The fact that $v$ corresponds to a minimum guarantees the absence of ``tadpoles,'' which are nothing but $\partial \Veff/\partial \phi (v) = 0$.

This simple picture is spoiled by several ``technical'' obstacles. 
First of all, it is very difficult to calculate $\Veff$ beyond the tree-level and to find an analytic expression for $v$.
Another well-known issue is the gauge dependencies \cite{Jackiw:1974cv,Nielsen:1975fs} of $\Veff$ (away from extrema) and the field value $v$ at its minimum. 
Due to this, various approximations for $v$ are on the market, 
which may lead to different PT series if expressed in terms of dimensionless running couplings and $m_\Phi^2$.

From the practical point of view, it is possible to avoid the explicit calculation of $\Veff$ by adjusting the definitions of the parameters 
(equivalently, the counterterms) to cancel tadpoles order by order (see Ref.~\cite{Sirlin:1985ux} for the on-shell formulation).

One can distinguish two, yet related, options (for a comprehensive discussion, see also Refs.~\cite{Actis:2006ra,Actis:2006rb,Actis:2006rc}).  
The first option is to write $\veff = \vtree + \Delta v$ and move terms due to $\Delta v$ to the interaction part of the SM Lagrangian in the broken phase,
so that all tree-level masses (c.f.\ Eq.\eqref{eq:mf_yf}) are proportional to $\vtree$, which satisfies the tree-level minimization condition, i.e., $\partial \Vtree(\vtree)/\partial \phi = 0$. 
The shift $\Delta v$ can be determined order by order in PT from the requirement that the tree-level tadpole for the neutral Higgs field $\phi$, 
which can be written as
\begin{equation}
	\delta {\mathcal L}_{\rm tad} = - \phi   t_{\rm tree} = -\phi   m_h^2   \Delta  v \left[ 1 + \frac{3}{2} \frac{\Delta v}{\vtree} + \frac{1}{2} 
	\left(\frac{\Delta v}{\vtree}\right)^2 \right], \qquad m_H^2 \equiv 2 \lambda \vtree^2,
	\label{eq:Ltad}
\end{equation}
cancels the loop-generated ones. 
In spite of the fact that the one-point function for the Higgs field is zero, in accordance with minimization condition, tadpoles 
in this scheme manifest themselves in every vertex in which we make the shift $\veff \to \vtree + \Delta v$. It turns out that such a kind of contributions
can be conveniently taken into account by considering one-particle-reducible diagrams in which a neutral-Higgs propagator 
is allowed to be terminated by a tadpole.
This approach was advocated by Fleischer and Jegerlehner (FJ) in Ref.~\cite{Fleischer:1980ub} (see also the recent discussion in Ref.~\cite{Denner:2016etu}.) 

Another prescription, the ``tadpole-free'' scheme, allows one to avoid the explicit introduction of $\Delta v$ and implicitly assumes that $v$ corresponds to a gauge-dependent field value at the minimum of $\Veff$ (see, e.g., Refs.~\cite{Degrassi:2012ry,Buttazzo:2013uya,Martin:2014cxa,Martin:2015lxa,Martin:2015rea,Martin:2016xsp}). 
In this case, the tree-level tadpole again precisely cancels the loop-generated ones, but Eq.~\eqref{eq:Ltad} is rewritten as
\begin{equation}
	\delta {\mathcal  L}_{\rm tad} = - \phi   t_{\rm tree} = -\phi   \veff   \left(\lambda \veff - m^2_{\Phi} \right),
\end{equation}
and there is no explicit contribution due to $\Delta v$.   
Due to this, the tree-level masses of the would-be Goldstone boson $\chi$ and
the Higgs boson $H$ are given by 
\begin{equation}
	m^2_{\chi} =  \lambda \veff^2 - m_{\Phi}^2 = 0 + \frac{t_{\rm tree}}{\veff}\,, \qquad m^2_H = 
	3 \lambda \veff^2 - m_{\Phi}^2  = 2 \lambda \veff^2 + \frac{t_{\rm tree}}{\veff}\,,
\label{eq:higgs_and_goldstone}
\end{equation}
in the broken phase, with $\veff^2\neq m^2_\Phi/\lambda$. 
Since it is a common choice to assume that all tree-level particle masses in the SM are proportional 
to a vev, the terms due to $t_{\rm tree}$ in Eqs.~\eqref{eq:higgs_and_goldstone} 
are moved from from the quadratic part of SM Lagrangian to the interaction part and are traded for loop-generated tadpoles.

It is worth pointing out here that $L$-loop contributions to the Higgs one-point functions considered 
in the two above-mentioned approaches, although being formally of the same loop level, are different, 
due to the fact that they are expressed in terms of different "tree-level" running masses. In addition, 
in the FJ scheme, Higgs tadpole insertions are allowed, while, in the ``tadpole-free'' scheme, only scalar masses are shifted due to tadpoles.

The advantage of the first option is an explicit control of the gauge dependence of the result, while, in the latter case, 
the Landau gauge, $\xi=0$, is usually chosen for the calculation of $\Veff$. In what follows, we routinely use the FJ tadpole scheme.

For the time being, we say nothing about the utilized regularization and our choice of renormalization scheme. 
This is done intentionally, since the reasoning is equally applicable if Eqs.~\eqref{eq:mf_yf}--\eqref{eq:higgs_and_goldstone} 
are written in terms of either bare or minimally renormalized parameters.
	For the FJ prescription, one defines a running vev $\vtree(\mu)^2\equiv m^2_{\Phi}(\mu)/\lambda(\mu)$, the RGE for which can be simply obtained from those of the 
	unbroken theory (see, e.g., Ref.~\cite{Bednyakov:2013cpa} and references cited therein). 
	In Refs.~\cite{Jegerlehner:2001fb,Jegerlehner:2002em,Jegerlehner:2002er,Jegerlehner:2012kn,Bezrukov:2012sa}, a related quantity, 
	$G_F^{\MSbar}(\mu) \equiv 1/(\sqrt2 v^2(\mu)$), is introduced (see below), and the RGEs are provided in terms 
	of running masses in the FJ tadpole scheme.  All running particle masses are gauge independent in this case 
	and are proportional to $\vtree(\mu)$ \cite{Jegerlehner:2002er}. In addition, the running Higgs mass is directly related to $m_\Phi^2$ of the unbroken Lagrangian, i.e., 
	$m_H^2 = 2 \lambda \vtree = 2 m_\Phi^2$.

One can also define a (gauge-dependent) running vev $\veff(\mu)$ obtained by minimization of the effective potential of the Higgs field, renormalized in the $\MSbar$ scheme at scale $\mu$, so that
$\veff(\mu) = \vtree(\mu) + \Delta v (\mu)$.  
The scale dependence of $\veff(\mu)$ is more involved than that of $\vtree$, but, in the Landau gauge, it is given by 
the Higgs field anomalous dimension (see the discussion in Refs.~\cite{Sperling:2013eva,Sperling:2013xqa}). 
The latter can also be calculated in the unbroken theory.

To summarize, we have discussed the following options to define a running quark
mass in the SM renormalized in the \MSbar{} scheme:
\begin{itemize}
	\item Gauge-independent running mass $\mb(\mu)$:
		\begin{align}
			\mb(\mu)  = \frac{y_b(\mu) \vtree(\mu)}{\sqrt2},&  \qquad \vtree(\mu)^2 \equiv \frac{m^2_{\Phi}(\mu)}{\lambda(\mu)}.
			\label{eq:mb_gi}
		\end{align}
	\item Gauge-dependent running mass $\tilde{m}_b(\mu)$:
		\begin{align}
			\tilde{m}_b(\mu)  = \frac{y_b(\mu) \veff(\mu)}{\sqrt2},& \qquad \veff(\mu): \left.\frac{\partial\Veff(\phi,\mu)}{\partial\phi}\right|_{\phi=\veff} = 0.
			\label{eq:mb_gd}
		\end{align}
\end{itemize}
	The anomalous dimensions $\gamma_m^b$ for both quantities, defined as
	\begin{align}
		\frac{d}{d \ln \mu} m = \gamma_m m, \qquad m \in \{ \mb,\tilde{m}_b \},
	\end{align}
		can be expressed as sums of the beta function $\beta_b$ for $y_b(\mu)$ and the anomalous dimensions $\gamma_b$ of the corresponding vevs $v$:
		\begin{align}
			\frac{d}{d \ln \mu} y_b = \beta_b y_b, \qquad \frac{d}{d \ln \mu} v = \gamma_v v,\qquad v \in \{\vtree, \veff\}. 
		\end{align}
	In the FJ case, we have \cite{Jegerlehner:2012kn}
\begin{align}
	\gamma_{\vtree} = \frac{1}{2} \left(\gamma_{m_{\Phi}^2}  - \frac{\beta_\lambda}{\lambda} \right),	
\end{align}
with $\gamma_{m_{\Phi}^2}\equiv d \ln m_{\Phi}^2/ \ln \mu$ and $\beta_\lambda \equiv d \lambda/d \ln \mu$. 
In the ``tadpole-free'' scheme, there is no such simple relation between the corresponding anomalous dimension and RG functions in a general $R_\xi$ gauge,
but, in Landau gauge, we have
\begin{align}
	\gamma_{\veff} = \gamma_{\Phi}, \qquad \gamma_{\Phi}  = - \frac{1}{2} \frac{d \ln Z_\Phi}{d \ln \mu}, \qquad \mathrm{(Landau~gauge!),}
\end{align}
	with $\gamma_{\Phi}$ being the anomalous dimension of the Higgs doublet $\Phi$, computed from the Higgs field 
	propagator renormalization constant $Z_\Phi$. 
	The difference in running between $\veff(\mu)$ and $\vtree(\mu)$ within the SM was studied numerically in Ref.~\cite{Bednyakov:2013cpa}.

	It is worth mentioning that, contrary to ``tadpole-free'' scheme,  
	all the gauge-fixing parameter dependences of calculable quantities 
	are explicit in the FJ approach. However, the corresponding expressions in the FJ scheme involve tadpole contributions,
	which typically scale like powers of $[M_t^4/(M_W^2 M_h^2)\sim 9]$ with $M_t$, $M_W$, and $M_H$ denoting the masses of the top quark, the $W$ boson, and the Higgs boson, respectively. 
	In the ``tadpole-free'' scheme, (most of) these dangerous terms are effectively absorbed in $\veff(\mu)$.

	As was stated earlier, we routinely use the FJ scheme to define running masses. Nevertheless, there is a way to improve the corresponding 
	PT series in a gauge-invariant way by trading $\vtree(\mu)$ for an ``on-shell'' vev, 
	\begin{align}
		\vF \equiv \left(\sqrt 2 G_F\right)^{-1/2} = 246.21965(6)~\GeV,
		\label{eq:vev_Fermi}
	\end{align}
	which, by definition, is related to a measured quantity, the Fermi constant extracted from muon decay, 
	$G_F\equiv G_\mu=1.1663787(6)\times 10^{-5}~\GeV^{-2}$ \cite{Agashe:2014kda}.
	In what follows, we treat $G_F$ as a non-renormalizable four-fermion coupling of the effective low-energy Fermi theory valid at scales 
	much less than the EW one (for a discussion of different definitions of the Fermi constant in the SM, see, e.g., Ref.~\cite{Jegerlehner:2012kn}). 
	The relation \eqref{eq:vev_Fermi} is motivated by the tree-level matching of the SM to the Fermi theory, 
	i.e., a $W$-boson exchange at low momentum transfer leads to
	\begin{align}
		\frac{G_F}{\sqrt 2} = \frac{g}{2 \sqrt 2} \times \frac{1}{M_W^2} \times \frac{g}{2 \sqrt 2} = \frac{1}{2 v^2}, \qquad M_W = \frac{g v}{2}, 
		\label{eq:Gf_tree_matching}
	\end{align}
		where the $W$-boson mass is proportional to the SU(2) gauge coupling $g$.
	Going beyond the tree-level approximation, one needs to perform a more sophisticated matching by comparing the QED-corrected and, at higher orders, also QCD-corrected muon lifetime in the EFT with the corresponding expression in the SM \cite{Awramik:2002vu}.
	The corrections to the tree-level matching in Eq.~\eqref{eq:Gf_tree_matching} are usually accumulated in the quantity $\Delta r$ \cite{Sirlin:1980nh},
	\begin{align}
		\frac{G_F}{\sqrt 2} = \frac{1}{2 v^2} \left(1 + \Delta r\right).
		\label{eq:Gf_loop_matching}
	\end{align}
	
	The Fermi constant can be treated as a Wilson coefficient of an effective non-renormalizable operator, which, in general, 
	can be scale dependent. We recall that Wilson coefficients are indeed
        scale dependent if the corresponding operator involves four external quarks \cite{Buras:1998raa}. 
	However, muon decay is described by an effective operator with external leptons,  
	$O_F = \left[ \bar \nu_\mu \gamma^\alpha ( 1 - \gamma_5 ) \mu \right] \left[ \bar e \gamma^\alpha ( 1 - \gamma_5) \nu_e\right]$,
	and the corresponding Wilson coefficient turns out to be scale independent 
	due to QED Ward--Takahashi identities (see the discussion in Ref.~\cite{Awramik:2002vu}), 
	i.e., $G_F$ can be treated as a scale-independent ``observable'' in the SM.

If the right-hand side of Eq.~\eqref{eq:Gf_loop_matching} is expressed in terms of $\MSbar$ parameters, one can invert it 
	in PT to express $v(\mu)$ in terms of $G_F$ and other (dimensionless) parameters. Again, $v$ can be either $\vtree(\mu)$ or $\veff(\mu)$.  
	It is easy to convince oneself that both the FJ and ``tadpole-free'' schemes should lead to the same PT series, if $v$ is traded for $G_F$ in  
	a consistent way.\footnote{One can also use a mixed renormalization scheme, for which the running masses in $\Delta r$ are rewritten
	in terms of pole ones.} Due to this, in our analysis, we make use of yet another definition of the running $b$-quark mass, 
	\begin{align}
		\mbY(\mu) \equiv \frac{y_b(\mu)\vF}{\sqrt 2}\,, \qquad \gamma_{\mbY} = \beta_b,
		\label{eq:mbY_definition}
	\end{align}
	discussed in Refs.~\cite{Jegerlehner:2003sp,Jegerlehner:2012kn,Kniehl:2014yia}.
	Since $\vF$ from Eq.~\eqref{eq:vev_Fermi} is scale independent, the anomalous dimension of $\mbY$ coincides with the Yukawa coupling beta function $\beta_b$.

\section{\label{sec:matching}Details of the matching procedure}

Let us now consider the relation between $\mbY(\mu)$ (or, equivalently, the $b$-quark Yukawa coupling)  
and the pole mass $M_b$ at the two-loop order, where we concentrate on EW corrections:
\begin{eqnarray}
	m_{b,Y}(\mu) & \equiv & \frac{y_b(\mu) \vF}{\sqrt 2}  = 
	M_b [ 1 + \delta_b(\mu) ], 
	\label{eq:y_pole_rel} \\
	\delta_b(\mu) & = & \sum\limits_{i+j=1}^{2}
	\alopi^i(\mu) 
	\asopi^j(\mu) 
	  \delta^{(b)}_{ij}(M_b,M,\mu)
	\label{eq:y_pole_rel_alpha} \\
	& = & 
	\sum\limits_{i+j=1}^{2}
	\alFopi^i 
	\asopi^j(\mu) 
	  \underline{\delta}^{(b)}_{ij}(M_b,M,\mu),
	\label{eq:y_pole_rel_alphaF} 
\end{eqnarray}
where $M\in\{M_W, M_Z, M_t, M_h\}$ collectively denotes the ``hard'' scales of the problem and $a_i(\mu)\equiv \alpha_i(\mu)/(4 \pi)$. In Eq.~\eqref{eq:y_pole_rel_alphaF}, instead of the running coupling $\al(\mu)$, a scale-independent coupling, 
$\alF\equiv \sqrt 2 G_F M_W^2\sin^2\theta_w/\pi = 1/132.233$ \cite{Agashe:2014kda}, where $\sin^2\theta_w= 1 - M_W^2/M_Z^2$, is used 
(see, e.g., Ref.~\cite{Kniehl:2015nwa} for a relation between $\al(\mu)$ and $\alF$).

We also need an (implicit) relation between the quark pole mass $M_b$ in $n_f=5$ QCD$\times$QED\footnote{We also consider three charged leptons in the spectrum.} and the running parameters 
	$\effmb(\mu)\equiv m_b^{(5)}(\mu)$, $\effasopi(\mu)\equiv a_s^{(5)}(\mu)$, and $\effalopi(\mu) \equiv a^{(5)}(\mu)$:
\begin{eqnarray}
	M_b & = & \effmb(\mu) \Big\{1 
		+ \effasopi   
		C_F \left( 4 + 3 \LMb \right) 
		+ \effalopi   
		Q_d^2 \left( 4 + 3 \LMb \right) 
		\nonumber\\
	    & + & 2   \effalopi \effasopi   
	    	C_F Q_d^2 \left[
	    \frac{121}{8} 
	    + 30 \zeta_2 
        + 8 I_3
	    + \frac{27}{2} \LMb + \frac{9}{2} \LLMb \right]
	    \nonumber\\
	    & + & \effalopi^2   
	    Q_d^2 (Q_e^2 + 2 Q_u^2)
	    	\left( 
	   	-\frac{71}{2} - 24 \zeta_2 - 26 \LMb - 6 \LLMb
		\right)
	\nonumber\\	
		& + & \effalopi^2   
		Q_d^4 
		\left(
		-\frac{1019}{8} + 30 \zeta_2 
        + 8 I_3
        - \frac{129}{2} \LMb - \frac{27}{2} \LLMb
		\right)
	\nonumber \\
	& + &  {\effasopi}^2   
	C_F \Big[ C_F \left( \frac{121}{8} + 30 \zeta_2 
    + 8 I_3
    \right) 
+ C_A \left( \frac{1111}{24} - 8 \zeta_2 
	- 4 I_3
\right)
\nonumber\\
&  & \phantom{\frac{\effas^2}{(4\pi)^2}}   - T_f \left( \left[\frac{71}{6} + 8 \zeta_2\right] n_f 
+ 12 n_h (1 -2 \zeta_2)  \right)
\nonumber\\
&  & \phantom{\frac{\effas^2}{(4\pi)^2}}   + \LMb \left( \frac{27}{2} C_F  + \frac{185}{6} C_A - \frac{26}{3} n_f 
T_f \right) 
\nonumber\\
& & 
	\phantom{\frac{\effas^2}{(4\pi)^2}}
					       + \LLMb \left( \frac{9}{2} C_F + \frac{11}{2} C_A  - 2 n_f 
				       T_f \right) \Big]
                        \Big\},
	\label{eq:polemass_qcd_qed}
\end{eqnarray}
where $Q_d=-1/3$, $Q_u = 2/3$, and $Q_e = -1$ are the electric charges of the SM fermions, $\LMb = \ln(\mu^2/M_b^2)$, and $I_3 = 3/2 \zeta_3 - 6 \zeta_2 \ln 2$. The QCD part for $n_l$ light flavors and $n_h$ heavy ones with a common mass, so that $n_f = n_l + n_h$, can be found, e.g., in Refs.~\cite{Gray:1990yh,Avdeev:1997sz,Fleischer:1998dw}.

	The pure-QED part can be obtained by the substitutions $C_A \to 0$, $C_F^2 \to Q_d^4$, $T_f n_f \to N_c (N_d Q_d^2 + N_u Q_u^2) + N_l Q_e^2$, and $T_f n_h\to Q_d^4 N_c n_h$, where $N_u = 2$, $N_d = 3$, $n_h = 1$, and $N_c = 3$. 

    The task is to relate $\effmb(\mu)$ to $\mbY(\mu)$ at a certain scale $\mu$ by introducing the so-called decoupling constants $\zeta(\mu)$: 
\begin{eqnarray}
	\effmb(\mu) & = & \mbY(\mu)   \zeta_{\mbY}(\mu), \quad \zeta_{\mbY}(\mu) = 1 + \delta \zeta_{\mbY}(\mu),
    \label{eq:mb_dec} \\
	\delta\zeta_{\mbY}(\mu) & = & \sum\limits_{i+j=1}^{2}
	\alopi^i 
	\asopi^j   
	\delta \zeta^{(b)}_{ij}(M,\mu) \label{eq:mb_dec_alpha} \\
	& = & \sum\limits_{i+j=1}^{2}
	\alFopi^i 
	\asopi^j   
	\deltaF \zeta^{(b)}_{ij}(M,\mu) \label{eq:mb_dec_alphaF}.
\end{eqnarray}
	The key feature of $\zeta_{\mbY}(\mu)$ is the absence of the dependence on the ``soft'' scale $M_b$ and 
    the absence of tadpole contributions. 
    The latter feature can be traced to the fact that, at the leading order, we have
    $\effmb(\mu) = \mbY(\mu)$ and not $\effmb(\mu) = \mb(\mu)\equiv y_b(\mu) \vtree(\mu)/\sqrt 2$ with a running, gauge-independent vev. 
    Nevertheless, it is worth mentioning that one can also use the latter definition for the extraction of the running Yukawa coupling $y_b(\mu)$. 
    This choice corresponds to restructuring the PT series in Eq.~\eqref{eq:mb_dec}.
    In spite of the fact that the decoupling constant $\effmb(\mu) = \mb(\mu)\zeta_{\mb}(\mu)$ in this case involve tadpole contributions, 
    the latter are canceled when $\mb(\mu)$ is divided by the running vev $v(\mu)$ expressed in terms of $\vF$ \cite{Kniehl:2015nwa}.
        However, our choice seems more natural, since QCD ``knows'' nothing about tadpoles and it is 
    tempting to absorb them in the effective mass parameter and not to put them into the decoupling constant.

    The perturbative expansion of $\zeta_{\mbY}(\mu)$ can be found order by order by
substituting  
    the pole mass of Eq.~\eqref{eq:polemass_qcd_qed} into Eq.~\eqref{eq:y_pole_rel} and expressing 
    $\effmb(\mu)$ in terms of $\mbY(\mu)$ via Eq.~\eqref{eq:mb_dec}.  
Expanding $\delta_{ij}^{(b)}$ in the small quantity $M_b$ and keeping only leading (logarithmic) terms,
	one obtains, at the one-loop order,
 $\delta \zeta^{(b)}_{01} = 0$, since there are no additional pure-QCD one-loop diagrams in the full SM, and 
	\begin{align}
	\delta \zeta^{(b)}_{10} & =  
    	- \frac{5}{18} 
        - \frac{1}{3} \LMZ
        + \frac{1}{\sin^2\theta_w} 
        \left[
        	\frac{41}{36} 
            + \frac{3}{4} \LMW 
            + \frac{1}{6}\LMZ 
        \right]
         -\frac{3}{8} \frac{\LMWZ}{\sin^4\theta_w}  
       \nonumber\\
       & +  
         \frac{1}{\sin^2 2\theta_w}
         \left[
         	  \frac{13}{9} 
            - \frac{1}{4 M_Z^2} ( M_t^2+ M_H^2 )
            + \frac{5}{6} \LMZ
            - \frac{3 M_t^2 }{2 M_Z^2} \LMT 
         \right]
         \nonumber \\
         & +  
           \frac{3}{8 \sin^2\theta_w}
           \left[
           	     \frac{M_H^2}{M_H^2 - M_W^2} \LMWH	
              -  \frac{M_t^2}{M_t^2 - M_W^2} 	
           	  -  \frac{M_t^2 M_W^2 }{(M_t^2 - M_W^2)^2} \LMWT	
           \right],
            \label{eq:xib10}
\end{align}
where $L_X\equiv \ln(\mu^2/M_X^2)$ and $L_{XY} = \ln(M_X^2/M_Y^2)$,
	since, in addition to photon exchange, we also have contributions involving the heavy EW gauge bosons and the Higgs boson in the full SM.
    The expression in Eq.~\eqref{eq:xib10} can be obtained from  the one-loop contribution to the ratio $\mbY(\mu)/M_b$  given in Eq.~(16) of Ref.~\cite{Kniehl:2014yia} by neglecting terms suppressed by powers of $M_b$ and subtracting pure QCD and QED terms.

    At the two-loop order, we have to take into account that the couplings of the $n_f=5$ QCD$\times$QCD effective theory should also be expressed in terms of more fundamental ones. 
    For the current work, it is sufficient to consider only one-loop decoupling relations (for results concerning $\alpha(\mu)$, see Refs.~\cite{Sirlin:1980nh,Erler:1998sy,Jegerlehner:2002em,Kniehl:2015nwa}):
\begin{align}
	\effas(\mu) & =  \as(\mu) \zeta_{\as}(\mu),\quad \zeta_{\as}(\mu) = 1 + \asopi \frac{4}{3} T_f \ln\frac{M_t^2}{\mu^2} + \ldots
    \label{eq:as_dec}\\
    \effal(\mu) & =  \al(\mu) \zeta_{\al}(\mu),\quad \zeta_{\al}(\mu) = 1 + \alopi  \underbrace{\left( \frac{2}{3} + \frac{4}{3} N_c Q_u^2 \ln \frac{M_t^2}{\mu^2} - 7 \ln \frac{M_W^2}{\mu^2}\right)}_{\delta \zeta^{(\alpha)}_{10}} + \ldots
    \label{eq:al_dec}
\end{align}
	The  decoupling constants given in Eqs.~\eqref{eq:as_dec} and \eqref{eq:al_dec} 
    can be easily obtained by expanding the required one-loop Green functions with external light particles in small external momenta and masses. 
    Only contributions with at least one heavy particle survive, and, in this
    simple case, no infrared divergences are generated. 
For the fine-structure constant, we also have to take into account the mixing of the photon with the $Z$ boson in the SM, so that we have
\begin{eqnarray}
	\delta\zeta^{(1)}_{\al}(\mu) & = & - \delta\zeta^{(1)}_{\gamma\gamma}(\mu) - \frac{\sin\theta_w}{\cos\theta_w} \delta\zeta^{(1)}_{Z\gamma}(\mu). 
\label{eq:al_dec_calc}
\end{eqnarray}
	Here, $\delta\zeta^{(1)}_{\gamma\gamma}(\mu)$ is found from the transverse part $i\Pi_{\gamma\gamma}(k^2)$ of the photon self-energy via the relation
\begin{eqnarray}
	\delta \zeta^{(1)}_{\gamma\gamma}(\mu) & = & \tilde \Pi^{(1)'}_{\gamma\gamma}(0),
\label{eq:alzeta_AA}
\end{eqnarray}
	in which the tilde is to indicate that one should only consider contribution from diagrams with at least one heavy line.
    For the mixing term in Eq.~\eqref{eq:al_dec_calc}, we have 
\begin{eqnarray}
	\delta \zeta^{(1)}_{Z\gamma}(\mu) = -\frac{2}{M_Z^2}\tilde \Pi^{(1)}_{Z\gamma}(0).
    \label{eq:alzeta_ZA}
\end{eqnarray}
	One can notice that Eqs.~\eqref{eq:al_dec_calc}---\eqref{eq:alzeta_ZA}
    resemble 
    expressions corresponding to the on-shell electric-charge renormalization at one loop (cf.\ Refs.~\cite{Sirlin:1985ux,Jegerlehner:2001fb,Jegerlehner:2002em,Awramik:2002vu})

    The result for the quark mass decoupling constant can be cross-checked by taking the derivative of Eq.~\eqref{eq:mb_dec} w.r.t.\ $\mu$, i.e.,
    \begin{align}
	    \gamma^b_m (\effas,\effal) & = \frac{1}{\zeta_{\mbY}}   \frac{d }{ d \ln \mu} 
	    	  \zeta_{\mbY}(\as,\al,M,\mu) + \beta_{y_b} (\as,\al,M,\mu).
	    \label{eq:gamma_beta_relation}
    \end{align}
    and expressing the effective-theory couplings, which appear on the left-hand side, in terms of the SM ones, $\al(\mu)$ and $\as(\mu)$.

    We expanded Eq.~\eqref{eq:gamma_beta_relation}
    in  $\as(\mu)$ and $\al(\mu)$ through the second order and proved that the relation indeed holds\footnote{
	    We can also trade the SM fine-structure constant $\al(\mu)$ for $\alF$ via Ref.~\cite{Kniehl:2015nwa}  
	    to prove that the scale dependence of $\deltaF\zeta^{(b)}_{ij}(\mu)$ is also reproduced.}	    
	    for $\delta \zeta^{(b)}_{11}(\mu) = \deltaF\zeta^{(b)}_{11}(\mu)$, $\delta\zeta^{(b)}_{20}(\mu)$, and 
	    $\deltaF\zeta^{(b)}_{20}(\mu)$ presented here.
	    The pure-QCD decoupling corrections for the running quark mass are known through the four-loop level \cite{Chetyrkin:1997un,Liu:2015fxa}.
	    For convenience, we present here the $\mathcal{O}(\as^2)$ term 
	    \begin{align}
		    \delta \zeta^{(b)}_{02}(\mu) & = C_F T_f \left( \frac{89}{18} 
		    - \frac{10}{3} \LMT 
		    + 2 \LMT^2 \right) 
		    \label{eq:zb02os}
	    \end{align}
	    and refer to Ref.~\cite{Chetyrkin:2000yt} for higher-order corrections. 

    The results 
    for the purely EW and mixed two-loop corrections can be cast into the following expressions\footnote{The same expansion holds for $\deltaF \zeta^{(b)}_{ij}(\mu)$, but with coefficients denoted by $\uXX{k}{ij}$.} with an auxiliary scale $\mu_0$:   
    \begin{align}
	    \delta\zeta^{(b)}_{ij}(\mu) = \left(\frac{M_t^2}{M_W^2 s_W^2} \right)^i \left( \XX{0}{ij} + \XX{1}{ij} \LL + \XX{2}{ij}\LLsq \right), \quad i+j = 2,	
	    \label{eq:Xkij_fYT}
    \end{align}
    	where the large ratio $M_t^2/(M_W^2\sin^2\theta_w) \simeq 20.8(2)$ was factored out. 
	We refrain from writing down a lengthy analytical result for the coefficients $\XX{k}{ij}$, but evaluate them at $\mu_0 = 175$~GeV and provide the following numerical formulas:  
	\begin{subequations}
	\begin{align}
	  \XX{2}{11} & = \frac{3}{2}, \\
          \XX{1}{11} & = 1.9647 - 0.0192\times\DeltaMT  - 0.0002\times \DeltaMW,  \\
		\XX{0}{11} & = -5.7665 - 0.0123\times \DeltaMT + 0.0015\times \DeltaMW - 0.0002\times \DeltaMZ, \\  
		\XX{2}{02} & = -0.365 + 0.001\times \DeltaMT, \\ 
		\XX{1}{02} &  = -0.329 + 0.016\times \DeltaMT, \\
		\XX{0}{02} & = -0.971 + 0.020\times \DeltaMT - 0.003\times \DeltaMW + 0.002\times \DeltaMH,    \\
		\uXX{2}{02} & = -0.389 + 0.001\times \DeltaMT, \\ 
		\uXX{1}{02} &  = -0.669 + 0.008 \times\DeltaMT, \\ 
		\uXX{0}{02} & = +0.569 + 0.012\times \DeltaMT, 
	\end{align}
		\label{eq:xi_num}
	\end{subequations}
    where $\Delta M_i \equiv (M_i - M_i^\mathrm{PDG})/\delta M^\mathrm{PDG}_i$ 
    with $M_i^\mathrm{PDG}$ and $\delta M_i^\mathrm{PDG}$ corresponding to the central value and experimental error for $M_i$ quoted by the PDG
\cite{Agashe:2014kda}. 
We have checked that Eq.~\eqref{eq:xi_num} reproduces the full analytic results\footnote{Available upon request from the authors.}  within the 3$\sigma$ region around the central values of the input parameters.

    The value of $\mbY(\mu)$ or, equivalently, the running Yukawa coupling in the SM can be found from $\effmb(\mu)$ by inverting the relation in Eq.~\eqref{eq:mb_dec} and
    expressing $\as(\mu)$ and $\al(\mu)$ in terms of the effective-theory couplings $\effas(\mu)$ and $\effal(\mu)$:

 \begin{align}
	 \mbY(\mu) & = \zeta^{-1}_{\mbY}(\mu) \effmb(\mu) = \effmb(\mu)  \sum\limits_{i+j=1}^{2}
	 \left(\effalopi\right)^i
	 \left(\effasopi\right)^j
	 \underline\delta\zeta^{(b)}_{ij}(\mu),
	 \label{eq:dec_inverted}
	 \\
	 \underline\delta\zeta^{(b)}_{10} & = -\delta\zeta^{(b)}_{10},\quad
	 \underline\delta\zeta^{(b)}_{11}  = -\delta\zeta^{(b)}_{11},\quad
	 \underline\delta\zeta^{(b)}_{02}  = -\delta\zeta^{(b)}_{02},\label{eq:inv_z10_z11_z02}\\
	 \underline\delta\zeta^{(b)}_{20} & = 
	 -\delta \zeta^{(b)}_{20}	+ \left( \delta \zeta^{(\al)}_{10} + \delta \zeta^{(b)}_{10} \right) \delta \zeta^{(b)}_{10}. 
	 \label{eq:inv_z20}
 \end{align}
 Another option is to use the scale-independent coupling $\alF$ in place of $\effal(\mu)$. In this case, one needs to replace $\delta\zeta^{(b)}_{20}(\mu)$ 
 by $\deltaF \zeta^{(b)}_{20}(\mu)$ in Eq.~\eqref{eq:inv_z20}
 and exclude the contributions from $\delta \zeta^{(\alpha)}_{10}(\mu)$, which originate in Eq.~\eqref{eq:dec_inverted} due to the conversion $\al(\mu)\to\effal(\mu)$. 

 	For illustration, let us present numerical values of the different corrections at some fixed scale, e.g., $\mu=M_Z$. In the case of the $\mbY - M_b$ relation, one obtains 
	\begin{align}
		\mbY(M_Z)  =  M_b \big(1 & 
					  - \underbrace{0.2682}_{\as} 
					  - \underbrace{0.0776}_{\as^2} 
					  - \underbrace{0.0330}_{\as^3} 
					  \nonumber \\
					  & 
					  - \underbrace{0.0101}_{\al} 
					  + \underbrace{0.0032}_{\al \as} 
					  + \underbrace{0.0003}_{\al^2} + \ldots \big),
		\label{eq:mbY_Mb_at_MZ}
	\end{align}
	while the $\mbY-\effmb$ relation yields
	\begin{align}
		\mbY(M_Z) = \effmb(M_Z) \big(1 & 
					       - \underbrace{0.00074}_{\as^2}	
					       - \underbrace{0.00023}_{\as^3}	
					       + \underbrace{0.00002}_{\as^4} + \Delta\zeta_b \big),	
		\label{eq:mbY_mbmb_at_MZ} \\
					       \Delta \zeta_b  = & 
					       - \underbrace{0.00838}_{\alF}
					       + \underbrace{0.00068}_{\alF \as}	
					       - \underbrace{0.00005}_{\alF^2}	+ \ldots 
					       \label{eq:mbY_mb_MZ_aF} \\
					       = & 
					       - \underbrace{0.00865}_{\al}
					       + \underbrace{0.00070}_{\al \as}	
					       + \underbrace{0.00029}_{\al^2}	+ \ldots 
	\end{align}
	From the comparison of Eqs.~\eqref{eq:mbY_Mb_at_MZ} and \eqref{eq:mbY_mbmb_at_MZ}, one can see that the decoupling corrections in Eq.~\eqref{eq:dec_inverted} are much smaller 
	than the pole-mass corrections in Eq.~\eqref{eq:y_pole_rel}, since the latter involve large logarithms, which are resummed in $\effmb(\mu)$ in the former case.
	As for the contribution due to EW interactions, the leading one-loop term dominates in Eq.~\eqref{eq:mbY_mbmb_at_MZ}, while the subleading mixed corrections of order $\mathcal{O}(\alF\al_s)$ tend to cancel 
	the two-loop $\mathcal{O}(\as^2)$ contribution. 

 	Equation~\eqref{eq:dec_inverted} is written for some fixed scale and is typically applied for $\mu\sim \mu_0$ close to the EW scale.
	The value of $\effmb(\mu_0)$ can be found from the known value $\mu_b \equiv \effmb(\effmb) = 4.18\pm0.03$~GeV\footnote{We use here the PDG value for conservative estimates, instead of the more precise value $\effmb(\effmb) = 4.136\pm0.016$~GeV \cite{Chetyrkin:2009fv}.} by solving the coupled RGEs 
	of the QCD$\times$QED effective theory,
	\begin{align}
		\frac{\effmb(\mu)}{\effmb(\mu_b)} & = \exp \left[~ \int\limits_{\mu_b}^{\mu} 
			\gamma_m^b\left[\effal(\mu'),\effas(\mu')\right] d \ln \mu'~\right] \equiv C_{\scriptstyle\mathrm{QCD}\times\mathrm{QED}}(\mu,\mu_b).
			\label{eq:mbmb_to_mbmu}
	\end{align}
	In pure QCD, the integration over $\ln \mu$ in Eq.~\eqref{eq:mbmb_to_mbmu} 
	can be traded for the integration over $\effas$. Due to this, the analogous factor $C_{\scriptstyle \mathrm{QCD}}(\mu,\mu_b)$ can be cast into 
	the form
	\begin{align}
		C_{\scriptstyle \mathrm{QCD}}(\mu,\mu_b) = \frac{c(\effas(\mu)/\pi)}{c(\effas(\mu_b)/\pi)},
	\end{align}
	with $c(x)$ given, e.g., in the recent Ref.~\cite{Baikov:2014qja}. 

	Collecting all the factors, the final formula for $\mbY(\mu)$ reads:
\begin{align}
	\mbY(\mu) = \mu_b C_{\mathrm{QCD}\times\mathrm{QED}}(\mu,\mu_b) \zeta^{-1}_{\mbY}(\mu). 
	\label{eq:mbY_master_formula}
\end{align}

\section{\label{sec:numerics}Numerical analysis of matching relations}

\begin{figure}[t]
		\centering
		\begin{subfigure}[b]{0.48\textwidth}
			{\hspace*{-5mm}\includegraphics[width=\textwidth,height=0.75\textwidth]{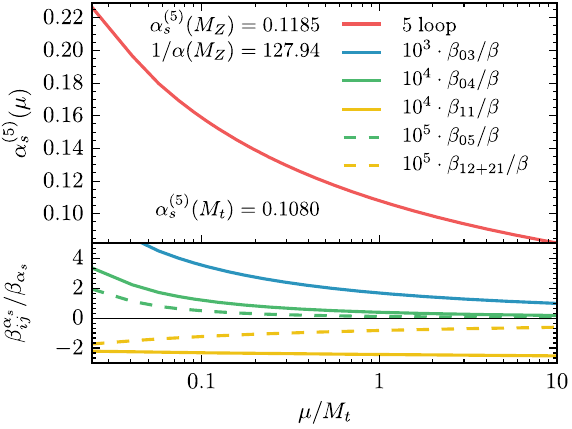}}
			\caption{$\beta_{\as} \equiv \sum\limits_{i+j=1}^3 \beta^{\as}_{ij} \alopi^i \asopi^j + \beta^{\as}_{04} \asopi^4 + \beta^{\as}_{05} \asopi^5$.}
			\label{fig:as_run_betas}
		\end{subfigure}
		~
		\begin{subfigure}[b]{0.48\textwidth}
			{\hspace*{-5mm}\includegraphics[width=\textwidth,height=0.75\textwidth]{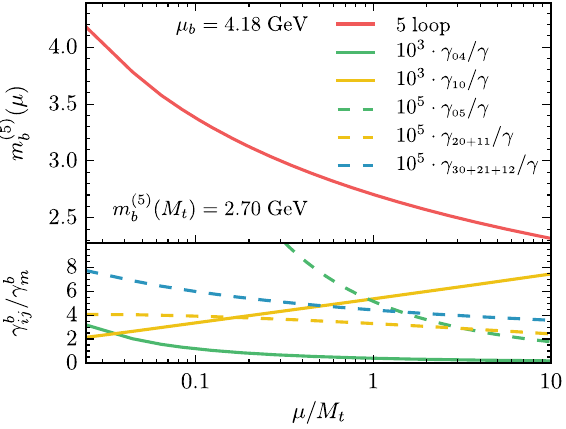}}
			\caption{$\gamma_m^b  \equiv \sum\limits_{i+j=1}^3 \gamma^{b}_{ij} \alopi^i \asopi^j + \gamma^{b}_{04} \asopi^4 + \gamma^{b}_{05} \asopi^5$.}
			\label{fig:mb_run_gamma}
		\end{subfigure}
		\hspace*{-1cm}
		\caption{
		The running of $\as$ and $\mb$ 
			in QCD$\times$QED with five active quark flavors
			obtained from the given input by means of five-loop RGEs. QED corrections are only included through the three-loop order.
			The effect of QED is negligible as compared to the uncertainty in the input parameters. 
			Nevertheless, in the case of $\alpha_s^{(5)}\equiv \effas$, the two-loop QED contribution to $\beta_{\as}$ is comparable with the four-loop QCD terms, while the three-loop corrections due to QED are of the same order 
			as the five-loop QCD result \cite{Baikov:2016tgj}.
			For the $b$-quark mass parameter $m_b^{(5)} \equiv \effmb$, the one-loop QED correction to $\gamma^b_m$ has the same order as the four-loop QCD term, while
			the five-loop contribution \cite{Baikov:2014qja} due to $\gamma^b_{05}$
			is much larger than the two- and three-loop QED corrections for $\mu \ll M_t$. For $\mu>M_t$, they become comparable. 
			It is also worth mentioning that, because of cancellations between terms due to $\gamma_{20}$ and $\gamma_{11}$, the two-loop QED corrections
			are even less than the three-loop ones.
		}
		\label{fig:qcdqed_run_betas}
	\end{figure}

	To begin with, we study the impact of additional QED correction to the running of $\effas(\mu)$ and  $\effmb(\mu)$. This running corresponds
	to the resummation of 
	logarithmically enhanced terms due to EW interactions.
	In Fig.~\ref{fig:qcdqed_run_betas}, we present the scale dependence of these quantities computed by means of the five-loop QCD RGE \cite{Baikov:2014qja,Baikov:2016tgj}
	accompanied by the three-loop QED corrections given in 
	\ref{sec:eft_RGE}. It turns out that the difference between three-loop and five-loop results are
	negligible when compared to experimental uncertainties in the boundary values.
	Nevertheless, for illustrative purpose, we provide the scale dependencies of the relative contributions to the five-loop strong-coupling beta function 
	and the quark mass anomalous dimension. 
	For the strong coupling, the two-loop QED contribution to $\beta_{\as}$ is comparable with the four-loop QCD terms, while the three-loop electromagnetic effects 
	compete with the five-loop pure-QCD contribution. As for the $b$-quark mass, the situation is similar, and the leading one-loop 
	QED corrections is of the same order as the four-loop pure-QCD terms. It is interesting to note that the corresponding 
	two-loop contributions are slightly less than the three-loop QED terms. This is due to a cancellation of $\mathcal{O}(\alpha^2)$ 
	and $\mathcal{O}(\alpha \alpha_s)$ corrections to $\gamma_{\effmb}$.

\begin{figure}[t]
		\centering
		\begin{subfigure}[b]{0.48\textwidth}
			{\hspace*{-5mm}\includegraphics[width=\textwidth,height=0.75\textwidth]{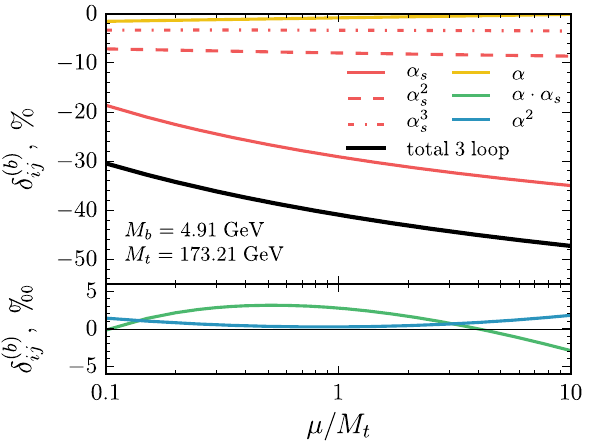}}
			\caption{Relation between $\mbY$ and $M_b$ \eqref{eq:y_pole_rel}.}
			\label{fig:mbY_cor_a}
		\end{subfigure}
		\begin{subfigure}[b]{0.48\textwidth}
			{\hspace*{-5mm}\includegraphics[width=\textwidth,height=0.75\textwidth]{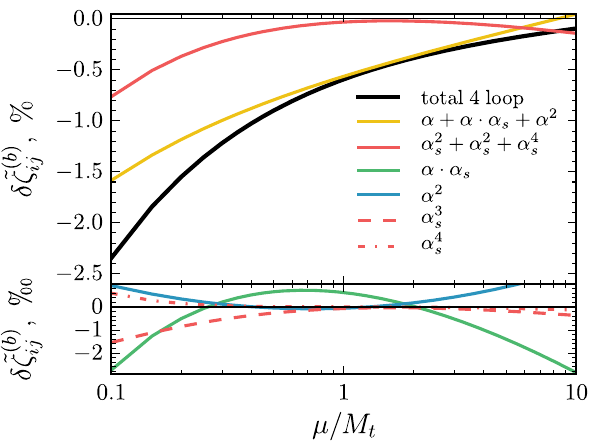}}
			\caption{Relation between $\mbY$ and $\effmb$ \eqref{eq:dec_inverted}.}
			\label{fig:mbY_cor_b}
		\end{subfigure}
		\hspace*{-1cm}
		\caption{The scale dependencies of different corrections to the relation between $\mbY$ and (a) the pole mass $M_b$ and (b) 
			the running mass $\effmb$. The reference scale is $M_t = 173.21$~GeV. 
		}
		\label{fig:DeltambYmbmb}
	\end{figure}

	Let us now perform a numerical analysis of the corrections to our matching formulas. In Fig.~\ref{fig:mbY_cor_a}, the scale dependencies of the different contributions to the relation in Eq.~ \eqref{eq:y_pole_rel} computed by means
of the program package \texttt{mr} \cite{Kniehl:2016enc} are presented.
	Note that the analytic expressions 
	for the two- and three-loop QCD corrections including finite top-quark
mass effects were taken from Ref.~\cite{Bekavac:2007tk}.
	The three-loop master integrals \cite{Bekavac:2009gz} were reevaluated numerically and by means of asymptotic expansion
	for the case of additional heavy quarks. Good agreement was found between numerical Mellin-Barnes integration and the lowest-order expansion.
	The corresponding expressions in the form of asymptotic series in the small parameter $z=M_b/M_t$ are given in 
	\ref{sec:zm_nm}.

	From Fig.~\ref{fig:mbY_cor_a} and Eq.~\eqref{eq:mbY_Mb_at_MZ}, one can see that pure-QCD contributions dominate the $\mbY-M_b$ relation. If one formally
	takes the value of the (total) three-loop term as an estimate of the theoretical uncertainty, the precision of the $\mbY-M_b$ matching in Eq.~\eqref{eq:y_pole_rel} 
	is currently limited to be a few percent due to the $\mathcal{O}(\as^3)$
contribution.  
	On the contrary, the PT series for $\mbY-\effmb$ in Eq.~\eqref{eq:dec_inverted} behaves much better. 
	Pure-QCD corrections involving $\ln(M_b/\mu)$ are resummed together with the QED ones,
	so that the relation is saturated by (one-loop) EW corrections, 
	which are about 1--2\%. Two-loop EW terms are approximately of the same order as
	three- and four-loop pure-QCD contributions. 
	If compared to the uncertainty of the input value of $\mu_b$, only the one-loop EW corrections turn out to be important 
	in the $\mbY-\effmb$ relation for the considered matching scales, 
	while dominant QCD corrections in the $\mbY-M_b$ relation can be resummed by means of the three-loop pure-QCD RGE. 

	Let us make one more comment about power-suppressed corrections of $\mathcal{O}(\mb/M)$ to the relation between $\mbY$ and $M_b$. 
	In our approach, we consistently neglect them. This also corresponds to dropping terms of the order of 
	$\alpha/(4\pi)\mb^2/M_W^2 \simeq y_b^2/(16 \pi^2)\sim 10^{-6}$ in Eq.~\eqref{eq:dec_inverted}.
	The estimated contribution is an order of magnitude less than the typical size of 
	the threshold corrections considered in this paper (c.f.\ Eq.~\eqref{eq:mbY_mb_MZ_aF}).
	Due to this, the inclusion of power-suppressed contributions is
	not necessary at the moment.

\begin{figure}[t]
		\centering
		\begin{subfigure}[b]{0.45\textwidth}
			{\includegraphics[width=\textwidth,height=0.75\textwidth]{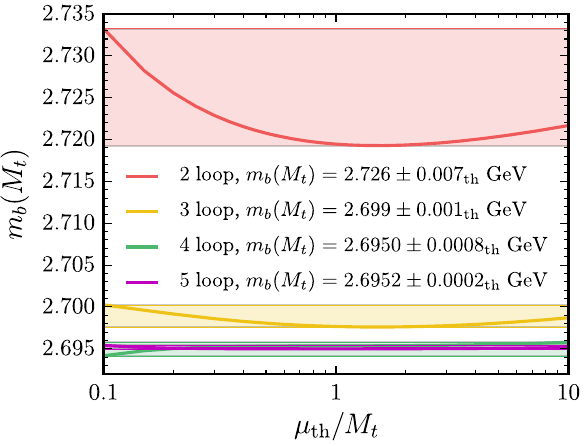}}
			\caption{QCD RGEs and threshold corrections.}
			\label{fig:qcd_a}
		\end{subfigure}
		~
		\begin{subfigure}[b]{0.45\textwidth}
			{\includegraphics[width=\textwidth,height=0.75\textwidth]{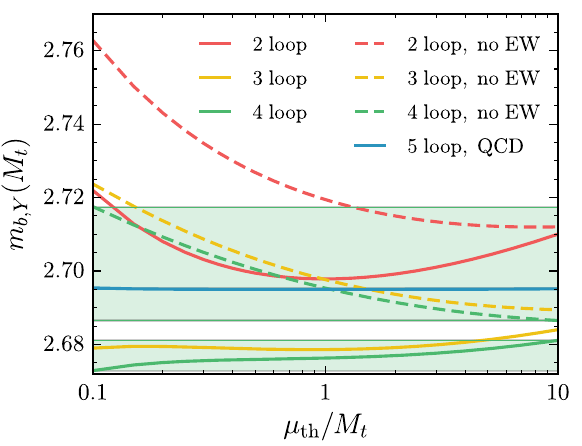}}
			\caption{SM RGEs and threshold corrections.}
			\label{fig:sm_b}
		\end{subfigure}
		\caption{The dependence on the matching scale $\mu_{\rm th}$ of the $b$-quark running mass parameter (a) in pure $n_f=6$ QCD, $\mb(M_t)$, and (b) 
			in the SM, $\mbY(M_t)$, at $L=2,3,4,5$ loops. 
			Pure-QCD threshold corrections are included through the $(L-1)$-loop level, and the corresponding values of $\mb(M_t)$ are indicated
			together with their theoretical uncertainties due to the $\mu_{\rm th}$ variation by a factor of ten. 
			In the case of the SM, EW and mixed contributions (collectively labeled EW) are taken 
			into account only through two loops. Four-loop contributions to the SM RGEs include pure-QCD corrections
			to the beta functions of the strong and quark Yukawa couplings together with recent results from Refs.~\cite{Bednyakov:2015ooa,Zoller:2015tha}.
			The necessity of EW threshold corrections in the SM can be deduced from the $\mu_{\rm th}$ scale dependence of the dashed curves, which lack the latter.
			In addition, the five-loop pure-QCD curve from Fig.~\ref{fig:qcd_a} is indicated. 
		}
		\label{fig:qedqcd_run_MT}
	\end{figure}

	Finally, we consider the dependence on the matching scale $\mu_{\rm th}$ of the running $b$-quark mass parameters in $n_f=6$ QCD, $\mb(M_t)$, 
	and the full SM, $\mbY(M_t)$, at a fixed scale $\mu=M_t$.   
	The running from $\mu_b$ to $\mu_{\rm th}$ is governed by the $n_f=5$ effective-theory RGEs, while the RG evolution from $\mu_{\rm th}$ to 
	$\mu=M_t$ is described by either QCD with active top quark or the full SM 
	(see Refs.~\cite{Mihaila:2012fm,Bednyakov:2012en,Chetyrkin:2013wya} for three-loop RGEs).
In Fig.~\ref{fig:qedqcd_run_MT}, one can see how the dependence is reduced 
	due to new higher-order terms both in the RGEs and the matching. 
	The $L$-loop RGEs are supplemented by $(L-1)$-loop threshold corrections in the pure-QCD case (see Fig.~\ref{fig:qcd_a}). 
	In Fig.~\ref{fig:qcd_a}, we also indicated our conservative estimates for the corresponding values of $\mb(M_t)$ 
	together with their theoretical uncertainty due to matching scale variation $0.1 \leq \mu_{\rm th}/M_t \leq 10$.

	In the SM, we lack three- and four-loop EW contributions to the $\mbY-\effmb$ relation. Moreover, four-loop EW corrections to the SM RGE are 
	only partially known in the literature \cite{Bednyakov:2015ooa,Zoller:2015tha}. Due to this, we restrict ourselves 
	in Fig.~\ref{fig:sm_b} to the four-loop order. The reduction of the matching-scale dependence is clearly visible when one
	goes from two to three loops in a self-consistent procedure, while the partial addition of four-loop (RG) terms does not improve the situation.
	From Fig.~\ref{fig:sm_b}, it is clear that, if we neglect the EW contribution in the matching relation as indicated by the dashed lines with the label ``no EW,'' 
	the dependence becomes more pronounced, thus, signifying the role of EW corrections in a consistent analysis. 

	In Fig.~\ref{fig:sm_b}, we also add the line corresponding to the four-loop pure-QCD result from Fig.~\ref{fig:qcd_a}. 
	Clearly, if one treats the QCD result as $\mbY(\mu)$ with neglected EW corrections, this overestimates $\mbY(\mu)$, and the shift
	is comparable with the experimental uncertainty in the input value of $\mu_b$, which is about 0.7\%. 
	Our final estimates for $\mbY(M_t)$ and the corresponding theoretical uncertainties are given by 
\begin{align}
	\mbY(M_t) & = 2.710 \pm 0.012_{\rm th}~\GeV \qquad \mbox{(2 loops)}, \nonumber \\
	\mbY(M_t) & = 2.681 \pm 0.003_{\rm th}~\GeV \qquad \mbox{(3 loops)},
	\label{eq:mbY_at_mt}
\end{align}
	from which the three-loop value of the corresponding Yukawa coupling can be easily obtained as
\begin{equation}
	y_b(M_t) = 0.01539\pm0.00002_{\rm th}.
	\label{eq:yb_at_mt}
\end{equation}
	On can see that, thanks to resummation of $\ln M_b/M_t$, the theoretical uncertainty is significantly reduced as compared to our previous analysis based on the $y_b-M_b$ relation
	\cite{Bednyakov:2015sca}.

	\section{\label{sec:conclusions}Conclusions}

	The $b$ quark plays a significant role in modern particle physics, and the precise knowledge of the corresponding 
	mass parameters is necessary for accurate theoretical predictions. 
	
	In this paper, we left aside low-energy problems related to confinement and considered high-energy, or short-distance, definitions
	of the $b$-quark mass. Given the value of the running $\MSbar$ mass in effective five-flavor QCD, 
	the EFT approach was used to relate it to the quantity of our interest, the running parameter in the SM,  $\mbY(\mu)$, 
	or, equivalently, the $b$-quark Yukawa coupling $y_b(\mu)$.
	We concentrated mainly on the two-loop EW corrections, which, although being suppressed with respect to the QCD ones,
	can play an appreciable role in precise analyses.

	We demonstrated how effective theories can be used to resum certain types of logarithmic corrections, e.g., $\ln(M_b/M)$, 
	which appear in the relation between $y_b(M)$ and the pole mass $M_b$.  
	Our analysis shows that the effect of QED logarithm resummation can be safely ignored at the moment, while 
	EW matching plays an important role in the estimation of the running parameter $y_b(\mu)$ at $\mu\geq M_Z$. 

	The obtained results for $y_b(\mu)$ mainly affect high-energy processes involving $b$ quarks, in which its (kinematic) mass can be neglected
	and only Yukawa interactions matter. As an example, we refer to the dominant Higgs decay mode $H\to \bar b b$ 
	(see Refs.~\cite{Bardin:1990zj,Kniehl:1991ze,Dabelstein:1991ky,Kataev:1997cq} and recent Ref.~\cite{Mihaila:2015lwa}
	for the EW corrections and Refs.~\cite{Gorishnii:1990zu,Chetyrkin:1997mb,Chetyrkin:1996sr,Chetyrkin:1997vj,Baikov:2005rw,Kniehl:1994ju,Chetyrkin:1996wr,Chetyrkin:1996ke} 
	for the corrections due to QCD). 

	\section*{Note Added}
	The published version of this paper contains a number of misprints (in $\delta \zeta^{(b)}_{10}$ and various $\beta^{\al}_{ij}$ and $\beta^{\as}_{ij}$)  and omissions (in $\gamma_{30}^b$), which have been corrected here. Fortunately, the computer code, which was used in the analysis, is free from these errors and the results and conclusions are not affected. 

\section*{Acknowledgments}
	The authors are grateful to M.Yu.~Kalmykov and A.I.~Onischenko for fruitful discussions. Multiple comments on the manuscript
	by M.Yu.~Kalmykov are also acknowledged. In addition, we would like to thank Steven Martin for pointing out misprints in the published version of the paper.
This work was supported in part by the German Federal Ministry for Education
and Research BMBF through Grant No.\ 05H15GUCC1, by the German Research
Foundation DFG through the Collaborative Research Centre No.\ SFB~676
{\it Particles, Strings and the Early Universe: the Structure of Matter and
  Space-Time}, by the Russian Foundation for Basic Research RFBR through Grant
No.~14-02-00494, and by the Heisenberg-Landau Programme.  

\appendix
\boldmath
\section{\label{sec:eft_RGE} RGEs in effective QCD$\times$QED}
\unboldmath

The RGEs for the effective-theory couplings 
are given by
(see also Refs.~\cite{Surguladze:1996hx,Erler:1998sy,Mihaila:2014caa})

\begin{align}
\mu^2 \frac{d \alpha_{i}}{d \mu^2} & = \alpha_i \beta^i, \quad \alpha_i \in  \left\{ \al, \as\right\}, \quad
\beta_i  = \sum\limits_{k,l=1}^3 \beta^{i}_{kl} \left(\alopi\right)^k 
\left(\asopi\right)^l + \ldots,
\label{eq:qcdxqed_rge_def}
\end{align}
\hspace*{-0.5cm}
\begin{align}
\beta^\al_{10} & = \frac{4}{3} \left[N_l Q_e^2   +   
      N_c \left(N_d Q_d^2 +  N_u Q_u^2\right)\right],  \qquad \beta^\al_{0j} \equiv 0,\quad j=1,...  \\
\beta^\al_{11} & = 4 C_F N_c \left[ N_d Q_d^2 +  N_u Q_u^2 \right], \\ 
\beta^\al_{20} & = 
4 \left[N_l Q_e^4  +   
      N_c \left(N_d Q_d^4 +  N_u Q_u^4\right)\right],  \\
\beta^\al_{30} & = 
 - \frac{44}{9}  N_c^2\left[N_d^2 Q_d^6 + N_u^2 Q_u^6 + N_u N_d \left( Q_u^2 + Q_d^2 \right) Q_u^2 Q_d^2 
 \right] \nonumber\\
 & 
 -\frac{44}{9} N_l Q_e^2 \left[  N_c \left[ Q_e^2 \left( N_u Q_u^2 + N_d Q_d^2\right) + N_u Q_u^4 + N_d Q_d^4  \right] + N_l Q_e^4\right]
 \nonumber\\
 & 
 -
2 \left[ N_c (N_u Q_u^6 + N_d Q_d^6) + N_l Q_e^6\right],
\\
\beta^\al_{21} & = - 4 C_F N_c \left[ N_u Q_u^4 + N_d Q_d^4 \right],
\\
\beta^\al_{12} & =  C_F N_c \left( N_u Q_u^2 + N_d Q_d^2\right)  \left[
\frac{133}{9} C_A - 2 C_F - \frac{44}{9} T_f (N_u + N_d) 
\right], 
\end{align}

\begin{align}
\beta^{\as}_{01} & = -\frac{11}{3} C_A + \frac{4}{3} T_f n_f 
, \qquad \beta^{\as}_{j0} \equiv 0, \quad j =1,...\\
\beta^{\as}_{11} & = 
4 T_F \left[ N_u Q_u^2 + N_d Q_d^2 \right], \\
\beta^{\as}_{02} & = - \frac{34}{3} C_A^2 + T_f n_f 
\left[ 4 C_F + \frac{20}{3} C_A \right], \\
\beta^{\as}_{03} & = - \frac{2857}{54} C_A^3 
+ \frac{1415}{27} C_A^2 T_f n_f 
+ \frac{205}{9} C_A C_F T_f n_f
- \frac{158}{27} C_A T_f^2 n_f^2
\nonumber \\
& - 2 C_F^2 T_f n_f
- \frac{44}{9} C_F T_f^2 n_f^2,\\
\beta^{\as}_{21} & = - \frac{44}{9} T_f   \left[ N_c (N_u Q_u^2 + N_d Q_d^2)^2 + N_l Q_e^2 (N_u Q_u^2 + N_d Q_d^2)  \right]\nonumber \\
& - 2 T_f (N_u Q_u^4 + N_d Q_d^4), \\
\beta^{\as}_{12} & =  4 T_f (2 C_A - C_F) (N_u Q_u^2 + N_d Q_d^2).
\end{align}

The anomalous dimension of the $b$-quark mass in QED$\times$QCD (for the pure-QCD part, see Refs.~\cite{Chetyrkin:1997dh,Vermaseren:1997fq}) can be cast into the form 
\begin{align}
\mu^2 \frac{ d m_b}{d \mu^2} & = \gamma^{b}_m m_b, \qquad
\gamma_m^b = - \sum\limits_{i+j=1}^{3} \gamma^b_{ij} 
\left( \alopi \right)^i   \left( \asopi \right)^j + \ldots, \\ 
\gamma^b_{01} & = 3 C_F, \qquad \gamma^b_{10} = 3 Q_d^2, \qquad \gamma^b_{11}  =  3 C_F Q_d^2, \\
\gamma^b_{02} & =  \frac{3}{2} C_F^2 + \frac{97}{6} C_F C_A - \frac{10}{3} C_F T_f n_f, \\
\gamma^b_{20} & = \frac{3}{2} Q_d^4 - \frac{10}{3} Q_d^2 \left[ N_c \left( N_u Q_u^2 + N_d Q_d^2 \right) + N_l Q_e^2\right] \\
\gamma^b_{03} & = - \frac{129}{4} C_F^2 C_A  + \frac{11413}{108} C_F C_A^2 + C_F C_A T_f n_f \left( - \frac{556}{27} - 48 \zeta_3 \right) \nonumber\\
& + \frac{129}{2} C_F^3  - \frac{140}{27} C_F T^2_f n^2_f + C_F^2 T_f n_f (- 45 + 48 \zeta_3) - C_F^2 T_f n_f, \\
\gamma^b_{12} & = - \frac{129}{4} C_F C_A Q_d^2 + 3    \frac{129}{2} C_F^2 Q_d^2  \nonumber \\
& - C_F T_f (N_u + N_d) Q_d^2 + C_F T_f (-45 + 48 \zeta_3) (N_d Q_d^2 + N_u Q_u^2)    \\
\gamma^b_{21} & =  3  \frac{129}{2} C_F Q_d^4 - C_F Q_d^2 \left[N_e Q_e^2 + N_c (N_u Q_u^2 + N_d Q_d^2) \right] \nonumber \\
	& + C_F Q_d^2 N_c (-45 + 48 \zeta_3) (N_d Q_d^2 + N_u Q_u^2) \\
\gamma^b_{30} & = \frac{129}{2} Q_d^6 - \frac{140}{27} Q_d^2 \left[N_e Q_e^2 + N_c (N_u Q_u^2 + N_d Q_d^2)\right]^2 \nonumber\\
	& -  Q_d^4 \left[ N_e Q_e^2 + N_c (N_u Q_u^2 + N_d Q_d^2)\right] \nonumber \\ 
	& + Q_d^2 (48 \zeta_3 - 45) \left[ N_e Q_e^4 + N_c (N_u Q_u^4 + N_d Q_d^4)\right] 
.
\end{align}

\section{\label{sec:zm_nm} Three-loop corrections to the pole mass of the $b$ quark in $n_f=6$ QCD}
\renewcommand\thefigure{\arabic{figure}}
	\begin{figure}[t]
			\centering{\includegraphics[width=0.55\textwidth]{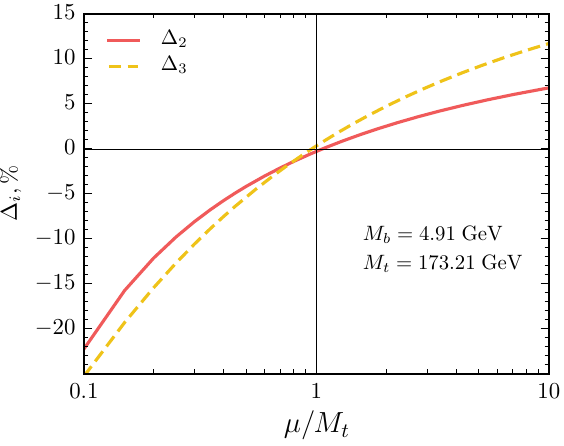}}
		\caption{The scale dependencies of the
		relative differences $\Delta_i\equiv 1 - \delta^{(b)}_{0i}(n_f=6)/\delta^{(b)}_{0i}(n_f=5)$
		of the pure-QCD coefficients of the $i$-loop contributions to the relation between the running and pole masses of the $b$ quark 
		in $n_f=5$ and $n_f=6$ QCD (cf.\ Eq.~\eqref{eq:y_pole_rel}). 
		}
		\label{fig:zm2}
	\end{figure}

Let us consider the relation between the pole mass $M_q$ and the running mass
$m_q(z,\mu)$ of a heavy quark in QCD with $n_l$ massless quarks, $n_h$ quarks with pole mass $M_q$, and $n_m$ quarks with pole mass $M_f$.
Defining $z\equiv M_q/M_f$, one can write the following relation: 
\begin{align}
  \label{eq:zm-def}
  \frac{m_q(z,\mu)}{M_q} 
  =1 & +\left(\delta_{\mathrm{QCD}}(\mu)\right)_{n_m=0} 
       + \left(\frac{\alpha_s}{4\pi}\right)^2 n_m X_{2,1}(M_q, z,\mu) \nonumber\\
     & +
       \left(\frac{\alpha_s}{4\pi}\right)^3\left[ n_m X_{3,1}(M_q, z,\mu) + n_m^2 X_{3,2}(M_q,z,\mu)\right].
\end{align}
	In the case of the $b$ quark in $n_f=6$ QCD, we have $n_l = 4$, $n_h = n_m = 1$, $M_q=M_b$, and $z=M_b/M_t$.

 The part independent of the heavy-quark masses can be found in
 Refs.~\cite{Chetyrkin:1999qi,Melnikov:2000qh,Kniehl:2015nwa}. The contributions from loops of
$n_m$ heavy quarks are contained in the coefficient $X_{2,1}$ at
two loops and in the coefficients $X_{3,1}$ and $X_{3,2}$ at three loops. The exact result for $X_{2,1}$ is available from Ref.~\cite{Gray:1990yh}. The expansions of $X_{3,1}$ and $X_{3,2}$ in the limit $z\to \infty$ are known from Ref.~\cite{Bekavac:2007tk}. Here,
we present results for $X_{3,1}$ and $X_{3,2}$ in the opposite limit $z\to 0$. The relation
between the masses is obtained from a general result \cite{Bekavac:2009gz},
in which the analytically known integrals were substituted and the unknown
$\mathcal{O}(\epsilon)$ parts of four master integrals were calculated by means of asymptotic expansion in the
large internal masses.

For convenience, we present here the two-loop result in expanded form,
\begin{align}
  \label{eq:zm-2l}
  X_{2,1}
  = C_F T_F & \left( -\frac{89}{18} + \frac{26}{3} L_{M} + 2
                             L_{M}^2 + \frac{52}{3}\ln(z) - 8\ln^2 (z)
                             \right. \nonumber\\
                           &\left. + z^2 \left(\frac{152}{75} - \frac{32}{15}
                             \ln(z)\right) + \mathcal{O}(z^4)\right),
\end{align}
together with the leading terms of the three-loop results,\footnote{The expansions up to the $\mathcal{O}(z^{10})$ terms
 can be found in an attachment to the arXiv version of this paper.}
\begin{align}
  \label{eq:zm-3l-nm2}
  X_{3,2}
  = & 
      C_F T_F^2 \left[\frac{3370}{243} -
      \frac{224}{9}\zeta_3 +
      \frac{496}{27} L_{M} - \frac{104}{9} L_{M}^2 -
      \frac{16}{9} L_{M}^3 \right.\nonumber\\
  + & 
      \left(\frac{992}{27} - \frac{416}{9}L_{M}\right)\ln(z) 
      - \left(\frac{416}{9} - \frac{64}{3}L_{M}\right) \ln^2 (z) + 
      \frac{256}{9}\ln^3 (z) \nonumber\\
  + & 
      \left.z^2 \left(\frac{368}{81} - \frac{1216}{225}L_{M} 
      - \left(\frac{2432}{225} - \frac{256}{45}L_{M}\right)\ln(z) 
      + \frac{512}{45}\ln^2 (z)\right) + \mathcal{O}(z^4) \right],
\end{align}

\begin{align}
  X_{3,1}
  = & 
      C_F^2 T_F\left[ \frac{547}{3} + \frac{88}{45}\pi^4 + \frac{32}{3}\pi^2
      \ln^2 2 - \frac{32}{3} \ln^4 2 - 256 a_4 -
      114 \zeta_3 \right.\nonumber\\
  + &  
      L_{M} \left(\frac{367}{6} + \frac{40}{3}\pi^2 -
      \frac{64}{3}\pi^2 \ln 2 - 16 \zeta_3\right) 
      - 26 L_{M}^2 - 6 L_{M}^3 \nonumber \\
  + & 
      \ln(z) \left(\frac{8}{3} + \frac{80}{3}\pi^2 -
      \frac{128}{3}\pi^2 \ln 2 - 52 L_{M}
      - 32\zeta_3\right)       \nonumber\\
  + & 
      24 L_{M} \ln^2 (z)  + z^2 \left(\frac{1001648}{30375} 
      + \frac{128}{135}\pi^2  - \frac{308}{9}\zeta_3 \right.\nonumber\\
  - &  
      \left.\left. \frac{152}{25}L_{M} -\left(\frac{26816}{2025} +
      \frac{32}{5}L_{M}\right) \ln(z) - \frac{128}{45} \ln^2 (z)\right) 
      \right] \nonumber\\ 
  + & 
      C_F T_F^2 \left[-\frac{5308}{243} ( n_h + n_l) +
      \frac{128}{9}\zeta_3(n_h + n_l) - 
      \left(40 n_h + 8 n_l \right.\right.\nonumber\\
  - & 
      \left.\frac{64}{9} n_h \pi^2 + \frac{32}{9} n_l
      \pi^2\right) L_{M} 
      -\frac{208}{9}(n_h + n_l)L_{M}^2 
      -\frac{32}{9}(n_h + n_l)L_{M}^3 \nonumber\\
  + &  
      \left(-80 n_h - 16 n_l + \frac{128}{9} n_h \pi^2 - \frac{64}{9} n_l
      \pi^2 - \frac{416}{9} (n_h + n_l) L_{M}\right)
      \ln(z) \nonumber\\
  + & 
      \frac{64}{3} (n_h + n_l) L_{M}
      \ln^2 (z) + \frac{128}{9}(n_h + n_l) \ln^3 (z) \nonumber\\
  + & 
      z^2 \left(-\frac{98624}{3375}n_h + \frac{33856}{3375}n_l +
      \frac{256}{135} n_h \pi^2 - \frac{128}{135} n_l \pi^2 
      -\frac{1216}{225}(n_h + n_l) L_{M} \right.\nonumber\\
  + & 
      \left.\left.\left(\frac{512}{25} n_h - \frac{128}{25} n_l + 
      \frac{256}{45}(n_h + n_l)L_{M}\right) \ln(z) 
      + \frac{256}{45} ( n_h + n_l)\ln^2 (z) \right)
      \right] \nonumber\\
  + & 
      C_F C_A T_F \left[-\frac{20083}{243} - \frac{62}{45}\pi^4 -
      \frac{16}{3} \pi^2 \ln^2 2 +  \frac{16}{3} \ln^4 2 + 
      128 a_4 + \frac{629}{9} \zeta_3 \right. \nonumber\\
  + & 
      L_{M}\left(\frac{1210}{9} - \frac{32}{9}\pi^2 +
      \frac{32}{3}\pi^2\ln 2 + 32\zeta_3\right) 
      + \frac{746}{9} L_{M}^2 + \frac{88}{9} L_{M}^3\nonumber\\   
  + & 
      \ln(z)\left(\frac{2612}{9} - \frac{64}{9}\pi^2 + \frac{64}{3}\pi^2
      \ln2 + \frac{1144}{9} L_{M} + 64 \zeta_3\right)\nonumber\\
  - & 
      \left(\frac{232}{3} + \frac{176}{3} L_{M}\right)\ln^2 (z) -
      \frac{352}{9}\ln^3 (z)  + z^2 \left(-\frac{996881}{60750} -
      \frac{25}{54} \pi^2 + \frac{124}{9}\zeta_3 \right. \nonumber\\
  + & 
      \left. \left. \frac{3344}{225} L_{M} +
      \left(\frac{115877}{4050} - \frac{704}{45} L_{M}\right) \ln(z) -
      \frac{454}{15} \ln^2 (z) \right) \right] + \mathcal{O}(z^4),
\end{align}
	where the abbreviations $L_{M}=\ln(\mu^2/M_q^2)$ and $a_4=\mathrm{Li}_4(1/2)$ are utilized.

	The effect due to the new terms is illustrated in Fig.~\ref{fig:zm2}, in which we present the relative differences of the two- and three-loop 
	pure-QCD coefficients $\delta^{(b)}_{0i}$ in Eq.~\eqref{eq:y_pole_rel} with and without the effect of the top quark as functions 
	of the renormalization scale. 
	We observe that, for low-mass scales $\mathcal{O}(M_b)$, the $n_f=5$ result underestimates the full $n_f=6$ corrections by more than 20\%, 
	while, for very large scales $\mathcal{O}(1~\mathrm{TeV})$, the effect is opposite both for the two-loop ($\Delta_2$) and 
	three-loop ($\Delta_3$) terms. One can see that, for scales of $\mathcal{O}(M_t)$, the difference is not so pronounced.\footnote{
	However, one should keep in mind that here we only compare coefficients of powers of $\as$.}


\begin{thebibliography}{10}
\expandafter\ifx\csname url\endcsname\relax
  \def\url#1{\texttt{#1}}\fi
\expandafter\ifx\csname urlprefix\endcsname\relax\def\urlprefix{URL }\fi
\expandafter\ifx\csname href\endcsname\relax
  \def\href#1#2{#2} \def\path#1{#1}\fi

\bibitem{Kobayashi:1973fv}
M.~Kobayashi, T.~Maskawa, {CP-violation in the renormalizable theory of weak interaction}, Prog. Theor. Phys. 49 (1973) 652--657,
\newblock \href {http://dx.doi.org/10.1143/PTP.49.652}
  {\path{doi:10.1143/PTP.49.652}}.

\bibitem{Herb:1977ek}
S.~W. Herb, et~al., {Observation of a dimuon resonance at
9.5~GeV in 400-GeV proton--nucleus collisions}, Phys. Rev. Lett. 39 (1977) 252--255,
\newblock \href {http://dx.doi.org/10.1103/PhysRevLett.39.252}
  {\path{doi:10.1103/PhysRevLett.39.252}}.

\bibitem{Bevan:2014iga}
A.~J. Bevan, et~al., {The physics of the $B$ factories}, Eur. Phys. J. C74 (2014)
  3026, 
\newblock \href {http://arxiv.org/abs/1406.6311} {\path{arXiv:1406.6311}},
  \href {http://dx.doi.org/10.1140/epjc/s10052-014-3026-9}
  {\path{doi:10.1140/epjc/s10052-014-3026-9}}.

\bibitem{Fleischer:2015mla}
R.~Fleischer, {Theoretical prospects for $B$ physics}, PoS FPCP2015 (2015) 002,
\newblock \href {http://arxiv.org/abs/1509.00601} {\path{arXiv:1509.00601}}.

\bibitem{Pich:1998xt}
A.~Pich, {Effective field theory: Course} (1998) 949--1049, 
\newblock \href {http://arxiv.org/abs/hep-ph/9806303} {\path{arXiv:hep-ph/9806303}}.

\bibitem{Grozin:2004yc}
A.~G. Grozin, {Heavy quark effective theory}, Springer Tracts Mod. Phys. 201
  (2004) 1--213,
\newblock \href {http://dx.doi.org/10.1007/b79301} {\path{doi:10.1007/b79301}}.

\bibitem{Bigi:1994em}
I.~I. Bigi, M.~A. Shifman, N.~Uraltsev, A.~Vainshtein, {The Pole mass of the
  heavy quark. Perturbation theory and beyond}, Phys.Rev. D50 (1994)
  2234--2246,
\newblock \href {http://arxiv.org/abs/hep-ph/9402360}
  {\path{arXiv:hep-ph/9402360}}, \href
  {http://dx.doi.org/10.1103/PhysRevD.50.2234}
  {\path{doi:10.1103/PhysRevD.50.2234}}.

\bibitem{Tarrach:1980up}
R.~Tarrach, {The pole mass in perturbative QCD}, Nucl. Phys. B183 (1981)
  384--396,
\newblock \href {http://dx.doi.org/10.1016/0550-3213(81)90140-1}
  {\path{doi:10.1016/0550-3213(81)90140-1}}.

\bibitem{Hoang:1998ng}
A.~H. Hoang, Z.~Ligeti, A.~V. Manohar, {$B$ decay and the $\Upsilon$ mass}, Phys.
  Rev. Lett. 82 (1999) 277--280,
\newblock \href {http://arxiv.org/abs/hep-ph/9809423}
  {\path{arXiv:hep-ph/9809423}}, \href
  {http://dx.doi.org/10.1103/PhysRevLett.82.277}
  {\path{doi:10.1103/PhysRevLett.82.277}}.

\bibitem{Beneke:1998rk}
M.~Beneke, {A quark mass definition adequate for threshold problems}, Phys.
  Lett. B434 (1998) 115--125, 
\newblock \href {http://arxiv.org/abs/hep-ph/9804241}
  {\path{arXiv:hep-ph/9804241}}, \href
  {http://dx.doi.org/10.1016/S0370-2693(98)00741-2}
  {\path{doi:10.1016/S0370-2693(98)00741-2}}.

\bibitem{Pineda:2001zq}
A.~Pineda, {Determination of the bottom quark mass from the $\Upsilon(1S)$
  system}, JHEP 06 (2001) 022,
\newblock \href {http://arxiv.org/abs/hep-ph/0105008}
  {\path{arXiv:hep-ph/0105008}}, \href
  {http://dx.doi.org/10.1088/1126-6708/2001/06/022}
  {\path{doi:10.1088/1126-6708/2001/06/022}}.

\bibitem{ElKhadra:2002wp}
A.~X. El-Khadra, M.~Luke, {The Mass of the $b$ quark}, Ann. Rev. Nucl. Part. Sci.
  52 (2002) 201--251,
\newblock \href {http://arxiv.org/abs/hep-ph/0208114}
  {\path{arXiv:hep-ph/0208114}}, \href
  {http://dx.doi.org/10.1146/annurev.nucl.52.050102.090710}
  {\path{doi:10.1146/annurev.nucl.52.050102.090710}}.

\bibitem{Peskin:2012we}
M.~E. Peskin, {Comparison of LHC and ILC capabilities for Higgs boson coupling measurements},
  \newblock \href {http://arxiv.org/abs/1207.2516} {\path{arXiv:1207.2516}}.

\bibitem{Klute:2013cx}
M.~Klute, R.~Lafaye, T.~Plehn, M.~Rauch, D.~Zerwas, {Measuring Higgs Couplings
  at a Linear Collider}, Europhys. Lett. 101 (2013) 51001,
\newblock \href {http://arxiv.org/abs/1301.1322} {\path{arXiv:1301.1322}},
  \href {http://dx.doi.org/10.1209/0295-5075/101/51001}
  {\path{doi:10.1209/0295-5075/101/51001}}.

\bibitem{Hempfling:1994ar}
R.~Hempfling, B.~A. Kniehl, {On the relation between the fermion pole mass
and $\overline{\mathrm{MS}}$ Yukawa coupling in the Standard Model}, Phys.Rev. D51 (1995) 1386--1394,
\newblock \href {http://arxiv.org/abs/hep-ph/9408313}
  {\path{arXiv:hep-ph/9408313}}, \href
  {http://dx.doi.org/10.1103/PhysRevD.51.1386}
  {\path{doi:10.1103/PhysRevD.51.1386}}.

\bibitem{Kniehl:2004hfa}
B.~A. Kniehl, J.~H. Piclum, M.~Steinhauser, {Relation between bottom-quark $\overline{\mathrm{MS}}$ Yukawa coupling and pole mass}, Nucl. Phys. B695 (2004) 199--216, 
\newblock \href {http://arxiv.org/abs/hep-ph/0406254}
  {\path{arXiv:hep-ph/0406254}}, \href
  {http://dx.doi.org/10.1016/j.nuclphysb.2004.06.036}
  {\path{doi:10.1016/j.nuclphysb.2004.06.036}}.

\bibitem{Kniehl:2014yia}
B.~A. Kniehl, O.~L. Veretin, {Two-loop electroweak threshold corrections to the
  bottom and top Yukawa couplings}, Nucl.Phys. B885 (2014) 459--480,
\newblock \href {http://arxiv.org/abs/1401.1844} {\path{arXiv:1401.1844}},
  \href {http://dx.doi.org/10.1016/j.nuclphysb.2015.02.012,
  10.1016/j.nuclphysb.2014.05.029; 10.1016/j.nuclphysb.2015.02.012,
  10.1016/j.nuclphysb.2014.05.029} {\path{doi:10.1016/j.nuclphysb.2015.02.012,
  10.1016/j.nuclphysb.2014.05.029; 10.1016/j.nuclphysb.2015.02.012,
  10.1016/j.nuclphysb.2014.05.029}}.

\bibitem{Marquard:2015qpa}
P.~Marquard, A.~V. Smirnov, V.~A. Smirnov, M.~Steinhauser, {Quark mass
  relations to four-loop order}, Phys.Rev.Lett. 114~(14) (2015) 142002,
\newblock \href {http://arxiv.org/abs/1502.01030} {\path{arXiv:1502.01030}},
  \href {http://dx.doi.org/10.1103/PhysRevLett.114.142002}
  {\path{doi:10.1103/PhysRevLett.114.142002}}.

\bibitem{Bekavac:2007tk}
S.~Bekavac, A.~Grozin, D.~Seidel, M.~Steinhauser, {Light quark mass effects in
  the on-shell renormalization constants}, JHEP 10 (2007) 006,
\newblock \href {http://arxiv.org/abs/0708.1729} {\path{arXiv:0708.1729}},
  \href {http://dx.doi.org/10.1088/1126-6708/2007/10/006}
  {\path{doi:10.1088/1126-6708/2007/10/006}}.

\bibitem{Bekavac:2009gz}
S.~Bekavac, A.~G. Grozin, D.~Seidel, V.~A. Smirnov, {Three-loop on-shell
  Feynman integrals with two masses}, Nucl. Phys. B819 (2009) 183--200,
\newblock \href {http://arxiv.org/abs/0903.4760} {\path{arXiv:0903.4760}},
  \href {http://dx.doi.org/10.1016/j.nuclphysb.2009.04.015}
  {\path{doi:10.1016/j.nuclphysb.2009.04.015}}.

\bibitem{Agashe:2014kda}
K.~Olive, et~al., {Review of Particle Physics}, Chin.Phys. C38 (2014) 090001,
\newblock \href {http://dx.doi.org/10.1088/1674-1137/38/9/090001}
  {\path{doi:10.1088/1674-1137/38/9/090001}}.

\bibitem{Coleman:1973jx}
S.~R. Coleman, E.~J. Weinberg, {Radiative corrections as the origin of spontaneous symmetry breaking}, Phys.Rev. D7 (1973) 1888--1910,
\newblock \href {http://dx.doi.org/10.1103/PhysRevD.7.1888}
  {\path{doi:10.1103/PhysRevD.7.1888}}.

\bibitem{Jackiw:1974cv}
R.~Jackiw, {Functional evaluation of the effective potential}, Phys.Rev. D9
  (1974) 1686,
\newblock \href {http://dx.doi.org/10.1103/PhysRevD.9.1686}
  {\path{doi:10.1103/PhysRevD.9.1686}}.

\bibitem{Nielsen:1975fs}
N.~Nielsen, {On the gauge dependence of spontaneous symmetry breaking in gauge theories}, Nucl.Phys. B101 (1975) 173,
\newblock \href {http://dx.doi.org/10.1016/0550-3213(75)90301-6}
  {\path{doi:10.1016/0550-3213(75)90301-6}}.

\bibitem{Sirlin:1985ux}
A.~Sirlin, R.~Zucchini, {Dependence of the Higgs coupling $\bar{h}_{\overline{\mathrm{MS}}}(M)$ on $m_H$ and the possible onset of new physics},
  Nucl.Phys. B266 (1986) 389,
\newblock \href {http://dx.doi.org/10.1016/0550-3213(86)90096-9}
  {\path{doi:10.1016/0550-3213(86)90096-9}}.

\bibitem{Actis:2006ra}
S.~Actis, A.~Ferroglia, M.~Passera, G.~Passarino, {Two-loop renormalization in the Standard Model. Part I: Prolegomena}, Nucl. Phys. B777 (2007) 1--34,
\newblock \href {http://arxiv.org/abs/hep-ph/0612122}
  {\path{arXiv:hep-ph/0612122}}, \href
  {http://dx.doi.org/10.1016/j.nuclphysb.2007.04.021}
  {\path{doi:10.1016/j.nuclphysb.2007.04.021}}.

\bibitem{Actis:2006rb}
S.~Actis, G.~Passarino, {Two-loop renormalization in the Standard Model. Part II:
Renormalization procedures and computational techniques}, Nucl.Phys. B777
  (2007) 35--99,
\newblock \href {http://arxiv.org/abs/hep-ph/0612123}
  {\path{arXiv:hep-ph/0612123}}, \href
  {http://dx.doi.org/10.1016/j.nuclphysb.2007.03.043}
  {\path{doi:10.1016/j.nuclphysb.2007.03.043}}.

\bibitem{Actis:2006rc}
S.~Actis, G.~Passarino, {Two-loop renormalization in the Standard Model. Part III:
Renormalization equations and their solutions}, Nucl.Phys. B777 (2007)
  100--156,
\newblock \href {http://arxiv.org/abs/hep-ph/0612124}
  {\path{arXiv:hep-ph/0612124}}, \href
  {http://dx.doi.org/10.1016/j.nuclphysb.2007.04.027}
  {\path{doi:10.1016/j.nuclphysb.2007.04.027}}.

\bibitem{Fleischer:1980ub}
J.~Fleischer, F.~Jegerlehner, {Radiative corrections to Higgs-boson decays in the
  Weinberg-Salam model}, Phys.Rev. D23 (1981) 2001--2026,
\newblock \href {http://dx.doi.org/10.1103/PhysRevD.23.2001}
  {\path{doi:10.1103/PhysRevD.23.2001}}.

\bibitem{Denner:2016etu}
A.~Denner, L.~Jenniches, J.-N. Lang, C.~Sturm, {Gauge-independent
	$\overline{\mathrm{MS}}$ renormalization in the 2HDM}, JHEP 09 (2016) 115,
\newblock \href {http://arxiv.org/abs/1607.07352} {\path{arXiv:1607.07352}},
  \href {http://dx.doi.org/10.1007/JHEP09(2016)115}
  {\path{doi:10.1007/JHEP09(2016)115}}.

\bibitem{Degrassi:2012ry}
G.~Degrassi, S.~Di~Vita, J.~Elias-Miro, J.~R. Espinosa, G.~F. Giudice, et~al.,
  {Higgs mass and vacuum stability in the Standard Model at NNLO}, JHEP 1208
  (2012) 098,
\newblock \href {http://arxiv.org/abs/1205.6497} {\path{arXiv:1205.6497}},
  \href {http://dx.doi.org/10.1007/JHEP08(2012)098}
  {\path{doi:10.1007/JHEP08(2012)098}}.

\bibitem{Buttazzo:2013uya}
D.~Buttazzo, G.~Degrassi, P.~P. Giardino, G.~F. Giudice, F.~Sala, et~al.,
  {Investigating the near-criticality of the Higgs boson}, JHEP 1312 (2013)
  089,
\newblock \href {http://arxiv.org/abs/1307.3536} {\path{arXiv:1307.3536}},
  \href {http://dx.doi.org/10.1007/JHEP12(2013)089}
  {\path{doi:10.1007/JHEP12(2013)089}}.

\bibitem{Martin:2014cxa}
S.~P. Martin, D.~G. Robertson, {Higgs boson mass in the Standard Model at
  two-loop order and beyond}, Phys.Rev. D90~(7) (2014) 073010,
\newblock \href {http://arxiv.org/abs/1407.4336} {\path{arXiv:1407.4336}},
  \href {http://dx.doi.org/10.1103/PhysRevD.90.073010}
  {\path{doi:10.1103/PhysRevD.90.073010}}.

\bibitem{Martin:2015lxa}
S.~P. Martin, {Pole mass of the $W$ boson at two-loop order in the pure $\overline{\mathrm{MS}}$ scheme}, Phys. Rev. D91~(11) (2015) 114003,
\newblock \href {http://arxiv.org/abs/1503.03782} {\path{arXiv:1503.03782}},
  \href {http://dx.doi.org/10.1103/PhysRevD.91.114003}
  {\path{doi:10.1103/PhysRevD.91.114003}}.

\bibitem{Martin:2015rea}
S.~P. Martin, {$Z$-boson pole mass at two-loop order in the pure $\overline{\mathrm{MS}}$ scheme}, Phys. Rev. D92~(1) (2015) 014026,
\newblock \href {http://arxiv.org/abs/1505.04833} {\path{arXiv:1505.04833}},
  \href {http://dx.doi.org/10.1103/PhysRevD.92.014026}
  {\path{doi:10.1103/PhysRevD.92.014026}}.

\bibitem{Martin:2016xsp}
S.~P. Martin, {Top-quark pole mass in the tadpole-free $\overline{\mathrm{MS}}$ scheme}, Phys. Rev. D93~(9) (2016) 094017,
\newblock \href {http://arxiv.org/abs/1604.01134} {\path{arXiv:1604.01134}},
  \href {http://dx.doi.org/10.1103/PhysRevD.93.094017}
  {\path{doi:10.1103/PhysRevD.93.094017}}.

\bibitem{Bednyakov:2013cpa}
A.~Bednyakov, A.~Pikelner, V.~Velizhanin, {Three-loop Higgs self-coupling
  beta-function in the Standard Model with complex Yukawa matrices}, Nucl.Phys.
  B879 (2014) 256--267,
\newblock \href {http://arxiv.org/abs/1310.3806} {\path{arXiv:1310.3806}},
  \href {http://dx.doi.org/10.1016/j.nuclphysb.2013.12.012}
  {\path{doi:10.1016/j.nuclphysb.2013.12.012}}.

\bibitem{Jegerlehner:2001fb}
F.~Jegerlehner, M.~Y. Kalmykov, O.~Veretin, {$\overline{\mathrm{MS}}$ vs.\ pole masses of gauge bosons: electroweak bosonic two-loop corrections}, Nucl.Phys. B641 (2002)
  285--326,
\newblock \href {http://arxiv.org/abs/hep-ph/0105304}
  {\path{arXiv:hep-ph/0105304}}, \href
  {http://dx.doi.org/10.1016/S0550-3213(02)00613-2}
  {\path{doi:10.1016/S0550-3213(02)00613-2}}.

\bibitem{Jegerlehner:2002em}
F.~Jegerlehner, M.~Y. Kalmykov, O.~Veretin, {$\overline{\mathrm{MS}}$
vs.\ pole masses of gauge bosons II: two-loop electroweak fermion corrections}, Nucl.Phys. B658 (2003)
  49--112,
\newblock \href {http://arxiv.org/abs/hep-ph/0212319}
  {\path{arXiv:hep-ph/0212319}}, \href
  {http://dx.doi.org/10.1016/S0550-3213(03)00177-9}
  {\path{doi:10.1016/S0550-3213(03)00177-9}}.

\bibitem{Jegerlehner:2002er}
F.~Jegerlehner, M.~Y. Kalmykov, O.~Veretin, {Full two loop electroweak
  corrections to the pole masses of gauge bosons}, Nucl.Phys.Proc.Suppl. 116
  (2003) 382--386,
\newblock \href {http://arxiv.org/abs/hep-ph/0212003}
  {\path{arXiv:hep-ph/0212003}}, \href
  {http://dx.doi.org/10.1016/S0920-5632(03)80204-9}
  {\path{doi:10.1016/S0920-5632(03)80204-9}}.

\bibitem{Jegerlehner:2012kn}
F.~Jegerlehner, M.~Y. Kalmykov, B.~A. Kniehl, {On the difference between the pole and the $\overline{\mathrm{MS}}$ masses of the
top quark at the electroweak scale},
  Phys.Lett. B722 (2013) 123--129,
\newblock \href {http://arxiv.org/abs/1212.4319} {\path{arXiv:1212.4319}},
  \href {http://dx.doi.org/10.1016/j.physletb.2013.04.012}
  {\path{doi:10.1016/j.physletb.2013.04.012}}.

\bibitem{Bezrukov:2012sa}
F.~Bezrukov, M.~Y. Kalmykov, B.~A. Kniehl, M.~Shaposhnikov, {Higgs boson mass and new physics}, JHEP 1210 (2012) 140,
\newblock \href {http://arxiv.org/abs/1205.2893} {\path{arXiv:1205.2893}},
  \href {http://dx.doi.org/10.1007/JHEP10(2012)140}
  {\path{doi:10.1007/JHEP10(2012)140}}.

\bibitem{Sperling:2013eva}
M.~Sperling, D.~St\"ockinger, A.~Voigt, {Renormalization of vacuum expectation
  values in spontaneously broken gauge theories}, JHEP 1307 (2013) 132,
\newblock \href {http://arxiv.org/abs/1305.1548} {\path{arXiv:1305.1548}},
  \href {http://dx.doi.org/10.1007/JHEP07(2013)132}
  {\path{doi:10.1007/JHEP07(2013)132}}.

\bibitem{Sperling:2013xqa}
M.~Sperling, D.~St\"ockinger, A.~Voigt, {Renormalization of vacuum expectation
  values in spontaneously broken gauge theories: Two-loop results},
  \newblock \href
  {http://arxiv.org/abs/1310.7629} {\path{arXiv:1310.7629}}.

\bibitem{Awramik:2002vu}
M.~Awramik, M.~Czakon, A.~Onishchenko, O.~Veretin, {Bosonic corrections to
  $\Delta r$ at the two loop level}, Phys.Rev. D68 (2003) 053004,
\newblock \href {http://arxiv.org/abs/hep-ph/0209084}
  {\path{arXiv:hep-ph/0209084}}, \href
  {http://dx.doi.org/10.1103/PhysRevD.68.053004}
  {\path{doi:10.1103/PhysRevD.68.053004}}.

\bibitem{Sirlin:1980nh}
A.~Sirlin, {Radiative corrections in the SU(2)${}_{L}$${}\times{}$U(1)
theory: A simple renormalization framework}, Phys.Rev. D22 (1980) 971--981,
\newblock \href {http://dx.doi.org/10.1103/PhysRevD.22.971}
  {\path{doi:10.1103/PhysRevD.22.971}}.

\bibitem{Buras:1998raa}
A.~J. Buras, \href{http://alice.cern.ch/format/showfull?sysnb=0282793}{{Weak
  Hamiltonian, CP violation and rare decays}}, in: {Probing the Standard Model
  of Particle Interactions. Proceedings, Summer School in Theoretical Physics,
  NATO Advanced Study Institute, 68th session, Les Houches, France, July
  28-September 5, 1997. Pt. 1, 2}, 1998, pp. 281--539,
\newblock \href {http://arxiv.org/abs/hep-ph/9806471}
  {\path{arXiv:hep-ph/9806471}}, \urlprefix\url{http://alice.cern.ch/format/showfull?sysnb=0282793}

\bibitem{Jegerlehner:2003sp}
F.~Jegerlehner, M.~Kalmykov, {$\mathcal{O}(\alpha\alpha_{s})$ relation between pole- and
$\overline{\mathrm{MS}}$-mass of the $t$-quark}, Acta Phys.Polon. B34 (2003)
  5335--5344,
\newblock \href {http://arxiv.org/abs/hep-ph/0310361}
  {\path{arXiv:hep-ph/0310361}}.

\bibitem{Kniehl:2015nwa}
B.~A. Kniehl, A.~F. Pikelner, O.~L. Veretin, {Two-loop electroweak threshold
  corrections in the Standard Model}, Nucl. Phys. B896 (2015) 19--51,
\newblock \href {http://arxiv.org/abs/1503.02138} {\path{arXiv:1503.02138}},
  \href {http://dx.doi.org/10.1016/j.nuclphysb.2015.04.010}
  {\path{doi:10.1016/j.nuclphysb.2015.04.010}}.

\bibitem{Gray:1990yh}
N.~Gray, D.~J. Broadhurst, W.~Grafe, K.~Schilcher, {Three-loop relation of quark $\overline{\mathrm{MS}}$ and pole masses}, Z.Phys. C48 (1990) 673--680,
\newblock \href {http://dx.doi.org/10.1007/BF01614703}
  {\path{doi:10.1007/BF01614703}}.

\bibitem{Avdeev:1997sz}
L.~V. Avdeev, M.~{\relax Yu}. Kalmykov, {Pole masses of quarks in dimensional
  reduction}, Nucl. Phys. B502 (1997) 419--435,
\newblock \href {http://arxiv.org/abs/hep-ph/9701308}
  {\path{arXiv:hep-ph/9701308}}, \href
  {http://dx.doi.org/10.1016/S0550-3213(97)00404-5}
  {\path{doi:10.1016/S0550-3213(97)00404-5}}.

\bibitem{Fleischer:1998dw}
J.~Fleischer, F.~Jegerlehner, O.~V. Tarasov, O.~L. Veretin, {Two loop QCD
  corrections of the massive fermion propagator}, Nucl. Phys. B539 (1999)
  671--690, [Erratum: Nucl. Phys.B571,511(2000)],
\newblock \href {http://arxiv.org/abs/hep-ph/9803493}
  {\path{arXiv:hep-ph/9803493}}, \href
  {http://dx.doi.org/10.1016/S0550-3213(99)00794-4,
  10.1016/S0550-3213(98)00705-6} {\path{doi:10.1016/S0550-3213(99)00794-4,
  10.1016/S0550-3213(98)00705-6}}.

\bibitem{Erler:1998sy}
J.~Erler, {Calculation of the QED coupling $\hat{\alpha} (M_Z)$ in the modified
minimal-subtraction scheme}, Phys. Rev. D59 (1999) 054008,
\newblock \href {http://arxiv.org/abs/hep-ph/9803453}
  {\path{arXiv:hep-ph/9803453}}, \href
  {http://dx.doi.org/10.1103/PhysRevD.59.054008}
  {\path{doi:10.1103/PhysRevD.59.054008}}.

\bibitem{Chetyrkin:1997un}
K.~Chetyrkin, B.~A. Kniehl, M.~Steinhauser, {Decoupling relations to
  $\mathcal{O} (\alpha_s^3)$ and their connection to low-energy theorems},
  Nucl.Phys. B510 (1998) 61--87,
\newblock \href {http://arxiv.org/abs/hep-ph/9708255}
  {\path{arXiv:hep-ph/9708255}}, \href
  {http://dx.doi.org/10.1016/S0550-3213(97)00649-4}
  {\path{doi:10.1016/S0550-3213(97)00649-4}}.

\bibitem{Liu:2015fxa}
T.~Liu, M.~Steinhauser, {Decoupling of heavy quarks at four loops and effective
  Higgs-fermion coupling}, Phys. Lett. B746 (2015) 330--334,
\newblock \href {http://arxiv.org/abs/1502.04719} {\path{arXiv:1502.04719}},
  \href {http://dx.doi.org/10.1016/j.physletb.2015.05.023}
  {\path{doi:10.1016/j.physletb.2015.05.023}}.

\bibitem{Chetyrkin:2000yt}
K.~Chetyrkin, J.~H. K\"uhn, M.~Steinhauser, {\texttt{RunDec}: a Mathematica package for running and decoupling of the
strong coupling and quark masses},
  Comput.Phys.Commun. 133 (2000) 43--65,
\newblock \href {http://arxiv.org/abs/hep-ph/0004189}
  {\path{arXiv:hep-ph/0004189}}, \href
  {http://dx.doi.org/10.1016/S0010-4655(00)00155-7}
  {\path{doi:10.1016/S0010-4655(00)00155-7}}.

\bibitem{Chetyrkin:2009fv}
K.~G. Chetyrkin, J.~H. K\"uhn, A.~Maier, P.~Maierh\"ofer, P.~Marquard,
  M.~Steinhauser, C.~Sturm, {Charm and bottom quark masses: An update}, Phys.
  Rev. D80 (2009) 074010,
\newblock \href {http://arxiv.org/abs/0907.2110} {\path{arXiv:0907.2110}},
  \href {http://dx.doi.org/10.1103/PhysRevD.80.074010}
  {\path{doi:10.1103/PhysRevD.80.074010}}.

\bibitem{Baikov:2014qja}
P.~A. Baikov, K.~G. Chetyrkin, J.~H. K\"uhn, {Quark mass and field anomalous dimensions to $\mathcal{O}(\alpha_{s}^{5})$}, JHEP 10 (2014) 076,
\newblock \href {http://arxiv.org/abs/1402.6611} {\path{arXiv:1402.6611}},
  \href {http://dx.doi.org/10.1007/JHEP10(2014)076}
  {\path{doi:10.1007/JHEP10(2014)076}}.

\bibitem{Baikov:2016tgj}
P.~A. Baikov, K.~G. Chetyrkin, J.~H. K\"uhn, {Five-loop running of the QCD coupling constant},
  \newblock \href {http://arxiv.org/abs/1606.08659}
  {\path{arXiv:1606.08659}}.

\bibitem{Kniehl:2016enc}
B.~A. Kniehl, A.~F. Pikelner, O.~L. Veretin, {\texttt{mr}: a C++ library for the matching and running of the
Standard Model parameters}, Comput. Phys. Commun.
  206 (2016) 84--96,
\newblock \href {http://arxiv.org/abs/1601.08143} {\path{arXiv:1601.08143}},
  \href {http://dx.doi.org/10.1016/j.cpc.2016.04.017}
  {\path{doi:10.1016/j.cpc.2016.04.017}}.

\bibitem{Mihaila:2012fm}
L.~N. Mihaila, J.~Salomon, M.~Steinhauser, {Gauge coupling beta functions in the Standard Model to three loops}, Phys.Rev.Lett. 108 (2012) 151602,
\newblock \href {http://arxiv.org/abs/1201.5868} {\path{arXiv:1201.5868}},
  \href {http://dx.doi.org/10.1103/PhysRevLett.108.151602}
  {\path{doi:10.1103/PhysRevLett.108.151602}}.

\bibitem{Bednyakov:2012en}
A.~Bednyakov, A.~Pikelner, V.~Velizhanin, {Yukawa coupling beta-functions in
  the Standard Model at three loops}, Phys.Lett. B722 (2013) 336--340,
\newblock \href {http://arxiv.org/abs/1212.6829} {\path{arXiv:1212.6829}},
  \href {http://dx.doi.org/10.1016/j.physletb.2013.04.038}
  {\path{doi:10.1016/j.physletb.2013.04.038}}.

\bibitem{Chetyrkin:2013wya}
K.~Chetyrkin, M.~Zoller, {$\beta$-function for the Higgs self-interaction in
  the Standard Model at three-loop level}, JHEP 1304 (2013) 091,
\newblock \href {http://arxiv.org/abs/1303.2890} {\path{arXiv:1303.2890}},
  \href {http://dx.doi.org/10.1007/JHEP04(2013)091, 10.1007/JHEP09(2013)155}
  {\path{doi:10.1007/JHEP04(2013)091, 10.1007/JHEP09(2013)155}}.

\bibitem{Bednyakov:2015ooa}
A.~V. Bednyakov, A.~F. Pikelner, {Four-loop strong coupling beta-function in
  the Standard Model},
  \newblock \href {http://arxiv.org/abs/1508.02680}
  {\path{arXiv:1508.02680}}, \href
  {http://dx.doi.org/10.1016/j.physletb.2016.09.007}
  {\path{doi:10.1016/j.physletb.2016.09.007}}.

\bibitem{Zoller:2015tha}
M.~F. Zoller, {Top-Yukawa effects on the $\beta$-function of the strong
  coupling in the SM at four-loop level}, JHEP 02 (2016) 095,
\newblock \href {http://arxiv.org/abs/1508.03624} {\path{arXiv:1508.03624}},
  \href {http://dx.doi.org/10.1007/JHEP02(2016)095}
  {\path{doi:10.1007/JHEP02(2016)095}}.

\bibitem{Bednyakov:2015sca}
A.~V. Bednyakov, B.~A. Kniehl, A.~F. Pikelner, O.~L. Veretin, {Stability of the electroweak vacuum: Gauge independence and advanced precision}, Phys. Rev.
  Lett. 115~(20) (2015) 201802,
\newblock \href {http://arxiv.org/abs/1507.08833} {\path{arXiv:1507.08833}},
  \href {http://dx.doi.org/10.1103/PhysRevLett.115.201802}
  {\path{doi:10.1103/PhysRevLett.115.201802}}.

\bibitem{Bardin:1990zj}
D.~{\relax Yu}. Bardin, B.~M. Vilensky, P.~K. Khristova, {Calculation of the
  Higgs boson decay width into fermion pairs}, Sov. J. Nucl. Phys. 53 (1991)
  152--158, [Yad. Fiz.53,240(1991)].

\bibitem{Kniehl:1991ze}
B.~A. Kniehl, {Radiative corrections for $H \to f \bar f (\gamma)$ in the
  standard model}, Nucl. Phys. B376 (1992) 3--28,
\newblock \href {http://dx.doi.org/10.1016/0550-3213(92)90065-J}
  {\path{doi:10.1016/0550-3213(92)90065-J}}.

\bibitem{Dabelstein:1991ky}
A.~Dabelstein, W.~Hollik, {Electroweak corrections to the fermionic decay width
  of the standard Higgs boson}, Z. Phys. C53 (1992) 507--516,
\newblock \href {http://dx.doi.org/10.1007/BF01625912}
  {\path{doi:10.1007/BF01625912}}.

\bibitem{Kataev:1997cq}
A.~L. Kataev, {Corrections of order $\mathcal{O} (\bar{\alpha}\bar{\alpha}_{s})$ and $\mathcal{O}
(\bar{\alpha}^{2})$ to the $\bar{b}b$-decay width of the neutral Higgs boson}, JETP Lett. 66 (1997) 327--330,
\newblock \href {http://arxiv.org/abs/hep-ph/9708292}
  {\path{arXiv:hep-ph/9708292}}, \href {http://dx.doi.org/10.1134/1.567516}
  {\path{doi:10.1134/1.567516}}.

\bibitem{Mihaila:2015lwa}
L.~Mihaila, B.~Schmidt, M.~Steinhauser, {$\Gamma(H\to b\bar{b})$ to order $\alpha\alpha_{s}$}, Phys. Lett. B751 (2015) 442--447,
\newblock \href {http://arxiv.org/abs/1509.02294} {\path{arXiv:1509.02294}},
  \href {http://dx.doi.org/10.1016/j.physletb.2015.10.078}
  {\path{doi:10.1016/j.physletb.2015.10.078}}.

\bibitem{Gorishnii:1990zu}
S.~G. Gorishnii, A.~L. Kataev, S.~A. Larin, L.~R. Surguladze, {Three-loop QCD correction to the correlator of the quark scalar currents and
$\Gamma_{\mathrm{tot}}(H^{0}\to \mbox{hadrons})$}, Mod. Phys. Lett. A5 (1990)
  2703--2712,
\newblock \href {http://dx.doi.org/10.1142/S0217732390003152}
  {\path{doi:10.1142/S0217732390003152}}.

\bibitem{Chetyrkin:1997mb}
K.~G. Chetyrkin, J.~H. K\"uhn, M.~Steinhauser, {Heavy quark current correlators
  to $\mathcal{O}(\alpha_s^2)$}, Nucl. Phys. B505 (1997) 40--64,
\newblock \href {http://arxiv.org/abs/hep-ph/9705254}
  {\path{arXiv:hep-ph/9705254}}, \href
  {http://dx.doi.org/10.1016/S0550-3213(97)00481-1}
  {\path{doi:10.1016/S0550-3213(97)00481-1}}.

\bibitem{Chetyrkin:1996sr}
K.~G. Chetyrkin, {Correlator of the quark scalar currents and $\Gamma_{\rm
  tot}(H\to{\rm hadrons})$ at $\mathcal{O} (\alpha_s^3)$ in pQCD}, Phys. Lett.
  B390 (1997) 309--317,
\newblock \href {http://arxiv.org/abs/hep-ph/9608318}
  {\path{arXiv:hep-ph/9608318}}, \href
  {http://dx.doi.org/10.1016/S0370-2693(96)01368-8}
  {\path{doi:10.1016/S0370-2693(96)01368-8}}.

\bibitem{Chetyrkin:1997vj}
K.~G. Chetyrkin, M.~Steinhauser, {Complete QCD corrections of order
  $\mathcal{O} (\alpha_s^3)$ to the hadronic Higgs decay}, Phys. Lett. B408
  (1997) 320--324,
\newblock \href {http://arxiv.org/abs/hep-ph/9706462}
  {\path{arXiv:hep-ph/9706462}}, \href
  {http://dx.doi.org/10.1016/S0370-2693(97)00779-X}
  {\path{doi:10.1016/S0370-2693(97)00779-X}}.

\bibitem{Baikov:2005rw}
P.~A. Baikov, K.~G. Chetyrkin, J.~H. K\"uhn, {Scalar correlator at $\mathcal{O}(\alpha_{s}^{4})$, Higgs boson decay into bottom quarks, and bounds on the light-quark masses}, Phys. Rev. Lett. 96 (2006) 012003,
\newblock \href {http://arxiv.org/abs/hep-ph/0511063}
  {\path{arXiv:hep-ph/0511063}}, \href
  {http://dx.doi.org/10.1103/PhysRevLett.96.012003}
  {\path{doi:10.1103/PhysRevLett.96.012003}}.

\bibitem{Kniehl:1994ju}
B.~A. Kniehl, M.~Spira, {Two-loop $\mathcal{O} (\alpha_{s} G_{F} m _{t}^{2})$ correction
to the $H \to b \bar{b}$ decay rate}, Nucl. Phys. B432 (1994) 39--48,
\newblock \href {http://arxiv.org/abs/hep-ph/9410319}
  {\path{arXiv:hep-ph/9410319}}, \href
  {http://dx.doi.org/10.1016/0550-3213(94)90592-4}
  {\path{doi:10.1016/0550-3213(94)90592-4}}.

\bibitem{Chetyrkin:1996wr}
K.~G. Chetyrkin, B.~A. Kniehl, M.~Steinhauser, {Virtual top-quark effects on the $H \to b\bar{b}$ decay at
next-to-leading order in QCD}, Phys. Rev. Lett.
  78 (1997) 594--597,
\newblock \href {http://arxiv.org/abs/hep-ph/9610456}
  {\path{arXiv:hep-ph/9610456}}, \href
  {http://dx.doi.org/10.1103/PhysRevLett.78.594}
  {\path{doi:10.1103/PhysRevLett.78.594}}.

\bibitem{Chetyrkin:1996ke}
K.~G. Chetyrkin, B.~A. Kniehl, M.~Steinhauser, {Three-loop $\mathcal{O} (\alpha^2_s G_F M_t^2)$ corrections to hadronic Higgs decays}, Nucl. Phys. B490 (1997)
  19--39,
\newblock \href {http://arxiv.org/abs/hep-ph/9701277}
  {\path{arXiv:hep-ph/9701277}}, \href
  {http://dx.doi.org/10.1016/S0550-3213(97)00051-5}
  {\path{doi:10.1016/S0550-3213(97)00051-5}}.

\bibitem{Surguladze:1996hx}
L.~R. Surguladze, {$\mathcal{O}(\alpha^{n} \alpha_{s}^{m})$ corrections in $e^{+}
e^{-}$ annihilation and $\tau$ decay}, 
  \newblock \href {http://arxiv.org/abs/hep-ph/9803211}
  {\path{arXiv:hep-ph/9803211}}.

\bibitem{Mihaila:2014caa}
L.~Mihaila, {Three-loop gauge beta function in non-simple gauge groups}, PoS
  RADCOR2013 (2013) 060.

\bibitem{Chetyrkin:1997dh}
K.~Chetyrkin, {Quark mass anomalous dimension to $\mathcal{O}(\alpha_s^4)$},
  Phys.Lett. B404 (1997) 161--165,
\newblock \href {http://arxiv.org/abs/hep-ph/9703278}
  {\path{arXiv:hep-ph/9703278}}, \href
  {http://dx.doi.org/10.1016/S0370-2693(97)00535-2}
  {\path{doi:10.1016/S0370-2693(97)00535-2}}.

\bibitem{Vermaseren:1997fq}
J.~A.~M. Vermaseren, S.~A. Larin, T.~van Ritbergen, {The 4-loop quark mass anomalous dimension and the invariant quark mass}, Phys. Lett. B405 (1997)
  327--333,
\newblock \href {http://arxiv.org/abs/hep-ph/9703284}
  {\path{arXiv:hep-ph/9703284}}, \href
  {http://dx.doi.org/10.1016/S0370-2693(97)00660-6}
  {\path{doi:10.1016/S0370-2693(97)00660-6}}.

\bibitem{Chetyrkin:1999qi}
K.~G. Chetyrkin, M.~Steinhauser, {The relation between the $\overline{\mathrm{MS}}$ and the on-shell quark mass at
order $\alpha_{s}^{3}$}, Nucl. Phys. B573 (2000) 617--651,
\newblock \href {http://arxiv.org/abs/hep-ph/9911434}
  {\path{arXiv:hep-ph/9911434}}, \href
  {http://dx.doi.org/10.1016/S0550-3213(99)00784-1}
  {\path{doi:10.1016/S0550-3213(99)00784-1}}.

\bibitem{Melnikov:2000qh}
K.~Melnikov, T.~v. Ritbergen, {The three-loop relation between the $\overline{\mathrm{MS}}$ and the pole quark masses}, Phys.Lett. B482 (2000) 99--108,
\newblock \href {http://arxiv.org/abs/hep-ph/9912391}
  {\path{arXiv:hep-ph/9912391}}, \href
  {http://dx.doi.org/10.1016/S0370-2693(00)00507-4}
  {\path{doi:10.1016/S0370-2693(00)00507-4}}.

\end{thebibliography}

\end{document}